\documentclass[11pt,a4paper]{article}

\usepackage{jheppub}
\usepackage{graphicx}
\usepackage{dcolumn}
\usepackage{bm}
\usepackage{amsmath}
\usepackage{braket}
\usepackage{slashed}
\usepackage{epstopdf}
\usepackage{placeins}
\usepackage{multirow}
\usepackage{makecell}
\usepackage{soul}
\usepackage{braket}
\usepackage[shortlabels]{enumitem}
\usepackage{placeins}
\usepackage[labelfont=bf, textfont=it, font=small]{caption}

\newcommand{\beq}{\begin{eqnarray}}
\newcommand{\eeq}{\end{eqnarray}}
\newcommand{\beqnn}{\begin{eqnarray*}}
\newcommand{\eeqnn}{\end{eqnarray*}}
\renewcommand{\thesection}{\arabic{section}}
\newcommand{\Tr}{\ensuremath{\mathrm{Tr}}}
\newcommand{\round}{\ensuremath{\mathrm{round}}}

\newcommand{\YM}{\ensuremath{\mathrm{YM}}}
\newcommand{\stag}{\ensuremath{\mathrm{stag}}}
\newcommand{\SP}{\ensuremath{\mathrm{SP}}}
\newcommand{\cool}{\ensuremath{\mathrm{cool}}}
\newcommand{\stout}{\ensuremath{\mathrm{stout}}}
\newcommand{\mc}{\ensuremath{\mathrm{mc}}}
\newcommand{\topo}{\ensuremath{\mathrm{topo}}}
\newcommand{\gluo}{\ensuremath{\mathrm{gluo}}}
\newcommand{\dof}{\ensuremath{\mathrm{dof}}}
\newcommand{\QCD}{\ensuremath{\mathrm{QCD}}}
\newcommand{\LQCD}{\ensuremath{\mathrm{LQCD}}}
\newcommand{\DIGA}{\ensuremath{\mathrm{DIGA}}}
\newcommand{\clov}{\ensuremath{\mathrm{clov}}}

\begin{document}
	
\title{Topological susceptibility of $N_f=2+1$ QCD from staggered fermions spectral projectors at high temperatures}

\author[a,b]{Andreas Athenodorou,}
\author[c,1]{Claudio Bonanno,\note{Corresponding author.}}
\author[a]{Claudio Bonati,}
\author[d]{Giuseppe Clemente,}
\author[a]{Francesco D'Angelo,}
\author[a]{Massimo D'Elia,}
\author[a]{Lorenzo Maio,}
\author[e]{Guido Martinelli,}
\author[f]{\\Francesco Sanfilippo}
\author[g,h,i]{and Antonino Todaro}

\affiliation[a]{Università di Pisa and INFN Sezione di Pisa, Largo B.~Pontecorvo 3, I-56127 Pisa, Italy}

\affiliation[b]{Computation-based Science and Technology Research Center, The Cyprus Institute, 20 Kavafi Str., Nicosia 2121, Cyprus}

\affiliation[c]{INFN Sezione di Firenze, Via G.~Sansone 1, I-50019 Sesto Fiorentino, Firenze, Italy}

\affiliation[d]{Deutsches Elektronen-Synchrotron (DESY), Platanenallee 6, 15738 Zeuthen, Germany}

\affiliation[e]{Dipartimento di Fisica and INFN Sezione di Roma ``La Sapienza'', Piazzale Aldo Moro 5, I-00185 Rome, Italy}

\affiliation[f]{INFN Sezione di Roma Tre, Via della Vasca Navale 84, I-00146 Rome, Italy}

\affiliation[g]{Department of Physics, University of Cyprus, P.O.~Box 20537, 1678 Nicosia, Cyprus}

\affiliation[h]{Faculty of Mathematics and Natural Sciences, University of Wuppertal, Wuppertal-42119, Germany}

\affiliation[i]{Dipartimento di Fisica, Università di Roma ``Tor Vergata'', Via della Ricerca Scientifica 1, I-00133 Rome, Italy}

\emailAdd{a.athenodorou@cyi.ac.cy}
\emailAdd{claudio.bonanno@fi.infn.it}
\emailAdd{claudio.bonati@unipi.it}
\emailAdd{giuseppe.clemente@desy.de}
\emailAdd{francesco.dangelo@phd.unipi.it}
\emailAdd{massimo.delia@unipi.it}
\emailAdd{lorenzo.maio@phd.unipi.it}
\emailAdd{guido.martinelli@roma1.infn.it}
\emailAdd{francesco.sanfilippo@infn.it}
\emailAdd{atodar01@ucy.ac.cy}

\abstract{

\noindent We compute the topological susceptibility of $N_f=2+1$ QCD with physical quark masses in the high-temperature phase, using numerical simulations of the theory
discretized on a space-time lattice. More precisely we estimate the topological
susceptibility for five temperatures in the range from $\sim200$~MeV up to
$\sim600$~MeV, adopting the spectral projectors definition of the topological charge 
based on the staggered Dirac operator. This strategy turns out to
be effective in reducing the large lattice artifacts which affect the standard
gluonic definition, making it possible to perform a reliable continuum extrapolation. Our results for the susceptibility in the explored temperature range are found to be partially in tension with previous determinations in the literature.}

\keywords{Lattice QCD, QCD Axion Phenomenology, $\mathrm{CP}$ Violation}

\arxivnumber{2208.08921}

\maketitle

\flushbottom

\section{Introduction}\label{sec:intro}

The study of the topological properties of QCD at high temperatures is of
utmost importance not only to provide better insight into the non-perturbative
regime of this theory, but also because of its phenomenological implications
for axion physics and cosmology. This justifies the interest in this
topic, which has been the subject of several recent Lattice QCD
investigations~\cite{Bonati:2015vqz, Frison:2016vuc,
Borsanyi:2016ksw, Petreczky:2016vrs, Bonati:2018blm, Lombardo:2020bvn}.

The axion is an hypothetical particle whose existence is predicted by the
Peccei-Quinn solution of the strong $\mathrm{CP}$-problem~\cite{Peccei:1977hh,
Peccei:1977ur, Wilczek:1977pj, Weinberg:1977ma}, 
that was also early recognized as a possible Dark Matter candidate.
Since the axion is directly coupled to the topological charge $Q$ defined by 
($F^{\mu\nu}$ is the QCD field strength)
\beq\label{eq:topocharge_continuum}
Q=\frac{1}{32 \pi^2} \varepsilon_{\mu\nu\rho\sigma}\int \Tr\{F^{\mu\nu}(x)F^{\rho \sigma}(x)\} d^4x\ ,
\eeq
the axion square mass is proportional to the topological susceptibility
$\chi=\langle Q^2\rangle /V$ ($V$ is the four-dimensional space-time volume).
More precisely $m_a^2=\chi/f_a^2$, where $f_a$ is the \emph{a priori} unknown axion
coupling scale.  An important peculiarity of axion physics is that it is
possible, modulo some general cosmological assumptions, to put an \emph{upper}
bound on the energy scale $f_a$~\cite{Preskill:1982cy, Abbott:1982af,
Dine:1982ah}. The specific value of this upper bound is fixed by the present
time Dark Matter abundance and by the temperature dependence of the axion
effective potential, i.e., mainly by the temperature dependence of the QCD topological
susceptibility~$\chi$. Since an upper bound on $f_a$ provides a lower bound for
the axion mass, the study of the QCD topological observables at finite
temperature provides an essential input for current and future experimental
axion searches (see~\cite{DiLuzio:2020wdo} for a recent review of the experimental bounds).

When the temperature is asymptotically high, one can compute $\chi(T)$ by
combining semiclassical methods and perturbation theory. 
Assuming instantons to be well separated and thus approximately not interacting with
each other (Dilute Instanton Gas Approximation, DIGA for short), and performing
a one loop computation in an instanton background, it is possible to obtain the 
result~\cite{Gross:1980br,Boccaletti:2020mxu}:
\beq\label{eq:DIGA_prediction_chi}
\chi(T) \propto T^{-c}\ ,\quad c=\frac{11}{3}N_c+\frac{1}{3}N_f-4\ ,
\eeq
where a logarithmic dependence of $\chi$ on the temperature has been implied and
$N_c,N_f$ are the number of colors and light flavors respectively. Being based on a
combination of semiclassical and perturbative approximations, this result is
expected to be trustworthy only for $T \gg \Lambda_{\QCD}\approx T_c$.
Nevertheless, due to the absence of more reliable computations, this result has
been routinely used to estimate the cosmological axion relic density needed to
constrain the axion coupling $f_a$. In performing this computation the
expression of $\chi(T)$ has however to be used starting from $T\gg T_c$ but
reaching temperatures of the order of $1\div 10$~GeV (see, e.g., Ref.~\cite{Wantz:2009it}), where non-perturbative deviations from the asymptotic
high-$T$ regime could be relevant. To avoid introducing systematic errors in
the computation of the axion coupling bound, it was thus suggested to use first
principles Lattice QCD results for $\chi(T)$ instead of the corresponding
DIGA expression~\cite{Berkowitz:2015aua}.

When exploring the high temperature regime of QCD with lattice
simulations, there are however several nontrivial numerical problems
that have to be faced. Among them the most notable are:
\begin{enumerate}[label=(\textit{\roman*})]
\itemsep0em 
\item\label{enum:prob_1} \emph{Rare $\vert Q \vert > 0$ events}\\
The topological susceptibility drops very rapidly with the temperature (DIGA predicts $c\approx 8$ for 3 light flavors, cf.~Eq.~\eqref{eq:DIGA_prediction_chi}),
thus the probability of visiting configurations with $Q \ne 0$ is rapidly
suppressed as the temperature is increased. This sampling problem has a physical origin and can be understood in terms of the vanishing of the
variance of the topological charge probability distribution $P(Q)$: since $\braket{Q^2}=\chi V \ll 1$ on affordable volumes, we have
$P(Q=0)\gg P(|Q|>0)$. From the numerical point of view, this implies the
necessity of collecting very large statistics in order to observe a sufficient
number of fluctuations above zero to reliably compute $\chi$.
\item\label{enum:prob_2} \emph{Explicit breaking of chiral symmetry and large lattice artifacts}\\
According to the \emph{index theorem}, the appearance of zero-modes in the
spectrum of the continuum massless Dirac operator $\slashed{D}$ is related to the
topological charge $Q$ of the background gauge field:
\beq\label{eq:index_theorem}
Q = n_+ - n_-\ ,
\eeq
where $n_+$ ($n_-$) is the number of zero modes with positive (negative)
chirality. For this reason, a finite volume configuration with non-zero
topological charge enters the continuum path integral with a weight that is
suppressed in the chiral limit as a power of the light quark mass. On the
lattice, however, if the sea quark discretization explicitly breaks the chiral
symmetry, no zero mode is present and no mode is chiral, being the smallest
eigenvalues shifted by cut-off effects. Such Would-Be Zero Modes
(WBZMs) make the suppression of $Q\ne 0$ configurations less efficient compared
to the continuum. As a consequence the numerical determination of $\chi$
is affected by large lattice artifacts, and its continuum extrapolation requires
particular care in the analysis of systematic uncertainties. In principle this
issue could be solved by simulating extremely fine lattice spacings, which is
in practice prevented by the \emph{topological critical slowing down} problem.
\item\label{enum:prob_3} \emph{Topological critical slowing down}\\
Local updating algorithms, such as the Rational Hybrid Monte Carlo
(RHMC)~\cite{Clark:2006fx,Clark:2006wp}, become less and less effective in changing the topological charge of
the configurations as the lattice spacing is decreased. While the critical
slowing down generically affects all observables, it is particularly severe for
the topological observables, for which the increase of the autocorrelation time
is consistent with an exponential in the inverse lattice
spacing~\cite{Alles:1996vn, DelDebbio:2004xh, Schaefer:2010hu, Luscher:2011kk, Bonati:2017woi, Bonanno:2018xtd}.
In practice, if the lattice spacing is small enough ergodicity is lost and the
whole Monte Carlo simulation remains trapped in the same topological sector;
as a matter of fact, this problem is also referred to as the ``freezing problem''. This makes
extremely difficult to reach, for the temperatures studied in this work,
lattice spacings of the order or smaller than $0.01$~fm.
\end{enumerate}

Overcoming these issues is mandatory to provide reliable lattice computations
of $\chi(T)$ at high temperatures, and several different strategies have been
proposed in literature to this end.
For instance, in Ref.~\cite{Borsanyi:2016ksw} (see also~\cite{Frison:2016vuc}),
the authors compute $\chi(T)$ bypassing issues~\ref{enum:prob_1}
and~\ref{enum:prob_3} simultaneously by restricting only to $Q=0$ and $\vert
Q\vert=1$, where neglecting contributions from higher-charge sectors is
justified on the basis of the DIGA  itself. The computational
problem~\ref{enum:prob_2}, related to the adoption of staggered fermions in the
lattice action, was instead addressed in Ref.~\cite{Borsanyi:2016ksw} by
reweighting configurations according to their expected continuum lowest eigenvalues of
$\slashed{D}$, thus trying to correct \emph{a posteriori} for the inaccurate sampling of chiral modes.
A crucial point to perform such a reweighting is the determination of
WBZMs, that are not so easily distinguishable from non-chiral modes without
some extra assumptions due to the explicit breaking of chiral symmetry on the
lattice.

The main goal of this work is to make progress towards an independent
determination of $\chi(T)$ in full QCD from the lattice without any extra
\emph{ad hoc} hypothesis. In particular, we will compute this quantity in $N_f=2+1$ QCD at the physical point combining the two following strategies:
\begin{enumerate}[label=(\textit{\alph*})]
\itemsep0em
\item\label{enum:strategy_1} \emph{Topological susceptibility from staggered
fermions spectral projectors }\\ 
In order to reduce the magnitude of lattice artifacts, we adopt a definition of
the topological susceptibility based on the Spectral Projectors (SP)
method~\cite{Luscher:2004fu, Giusti:2008vb, Luscher:2010ik, Cichy:2015jra,
Alexandrou:2017bzk}, introduced for staggered fermions in
Ref.~\cite{Bonanno:2019xhg}. This discretization, based on a lattice version of the
index theorem, consists of defining $Q$ as the sum of the chiralities of all
the modes lying below a certain threshold $M$. The SP topological susceptibility is a theoretically well-defined quantity, as it can be shown to converge to the correct continuum
limit, and the choice of $M$ can be used to reduce the lattice
artefacts with respect to the standard gluonic definition of the topological
charge. Our approach will allow us to have a better control on the
systematic uncertainties related to the continuum extrapolation of $\chi$ already in the lattice spacing range employed in typical simulations, thus
alleviating the necessity for extremely fine lattice spacings to reduce lattice artifacts, whose use would require a specific strategy to deal with the issue~\ref{enum:prob_3}.
\item\label{enum:strategy_2} \emph{Multicanonical algorithm}\\
The multicanonical algorithm~\cite{Berg:1992qua} was introduced to
study strong first order phase transitions, and was adapted
in~\cite{Bonati:2017woi, Jahn:2018dke, Bonati:2018blm, Bonanno:2022dru} to deal with the problem
of the dominance of the $Q=0$ sector at high temperatures
(issue~\ref{enum:prob_1} above). The idea is to add to the lattice action a
fixed (as opposed to the case of metadynamics~\cite{Laio:2015era}) bias
topological potential in order to enhance the probability of visiting
suppressed topological sectors, thus effectively enhancing the fluctuations of
$Q$ during the Monte Carlo evolution. Expectation values with respect to the
original path-integral distribution are then exactly reproduced by using
standard reweighting.  With respect to the standard approach, in this way a
much better statistical accuracy is achieved in the determination of the
relative weights of the topological sectors.
\end{enumerate}

We here anticipate that our results for $\chi(T)$ obtained from spectral projectors show some difference with respect to those reported in Ref.~\cite{Borsanyi:2016ksw}, as we observe a $\sim 2-3$ standard deviation tension in a temperature range between $300$ and $400$~MeV. We also observe, in the same temperature range, a milder $\sim2-2.5$ standard deviation tension with respect to the determinations of Ref.~\cite{Petreczky:2016vrs}, which however had to be mass-extrapolated to be compared with our results, as will be discussed in the following.

This paper is organized as follows: in Sec.~\ref{sec:num_setup} we present our
numerical setup, discussing in particular the spectral discretization
of the topological susceptibility~\ref{enum:strategy_1} and our
implementation of the multicanonical algorithm~\ref{enum:strategy_2}.
In Sec.~\ref{sec:results} we present continuum-extrapolated results for the
topological susceptibility in $N_f=2+1$ QCD at the physical point. First, we
apply the SP approach to the $T=0$ case, which is used as a test-bed
to compare this method with the standard gluonic approach. Then, we
adopt the same SP method in combination with the multicanonical
approach to compute the topological susceptibility as a function of the
temperature in the range $200\,\text{MeV} \lesssim T \lesssim 600\,\text{MeV}$.
Our results are then compared with other determinations in the literature and
with DIGA predictions. Finally, in Sec.~\ref{sec:conclusions} we draw our
conclusions and discuss future outlooks of this work.

\section{Numerical setup}\label{sec:num_setup}

\subsection{Lattice action}

We discretize $N_f=2+1$ QCD on a $N_s^3\times N_t$ lattice adopting rooted
stouted staggered fermions for the quark sector and the tree-level Symanzik
improved gauge action for the gluon sector. The partition function $Z_{\LQCD}$ is thus given by
\beq\label{eq:partition_function}
Z_{\LQCD}=\int [dU] e^{-S^{(L)}_\YM[U]} \det\left\{\mathcal{M}^{(\stag)}_{l}[U]\right\}^{\frac{1}{2}} 
\det\left\{\mathcal{M}^{(\stag)}_{s}[U]\right\}^{\frac{1}{4}}\ ,
\eeq
where $u=d\equiv l$ and $s$ denote the two mass-degenerate light quarks and the
strange quark respectively. The staggered Dirac operator $D_{\stag}$ is
defined by using the gauge links $U_{\mu}^{(2)}$, obtained by applying to the
gauge configuration $n_{\stout}=2$ levels of isotropic stout
smearing~\cite{Morningstar:2003gk} with $\rho_{\stout}=0.15$, thus:
\begin{gather}
\mathcal{M}^{(\stag)}_f[U] \equiv D_{\stag}[U^{(2)}] + \hat{m}_f, \qquad \hat{m}_f \equiv am_f\ ,\nonumber\\
\label{eq:stag_operator}
D_{\stag}[U^{(2)}] = \sum_{\mu=1}^{4}\eta_{\mu}(x)\left( U^{(2)}_{\mu}(x) \delta_{x,y-\hat{\mu}} - {U_{\mu}^{(2)}}^{\dagger}(x-\hat{\mu}) \delta_{x,y+\hat{\mu}} \right)\ ,\\ 
\eta_{\mu}(x) = (-1)^{x_1+\dots+x_{\mu-1}}\ . \nonumber
\end{gather}
The tree-level Symanzik-improved Wilson action is instead expressed in terms of
the non-stouted gauge links:
\beq
S_{\YM}^{(L)}[U] = - \frac{\beta}{3} \sum_{x, \mu \ne \nu} \left\{ \frac{5}{6}\Re\Tr\left[\Pi_{\mu\nu}^{(1\times1)}(x)\right] 
- \frac{1}{12}\Re\Tr\left[\Pi_{\mu\nu}^{(1\times2)}(x)\right]\right\}\ ,
\eeq
where $\Pi^{(n\times m)}_{\mu\nu}(x)$ is the $n\times m$ Wilson loop.

The bare parameters $\beta$, $\hat{m}_s$ and $\hat{m}_u=\hat{m}_d \equiv
\hat{m}_l$ are tuned in order to move on a Line of Constant Physics (LCP) corresponding to the physical values of the pion mass $m_{\pi} \simeq 135$~MeV and of the ratio
$\hat{m}_s / \hat{m}_l = m_s/m_l \simeq 28.15$ \cite{Aoki:2009sc, Borsanyi:2010cj, Borsanyi:2013bia}.

\subsection{Topological charge discretizations}

Topological charge definitions can be divided into two broad groups: gluonic
and fermionic ones. 
The simplest gluonic definition is the clover one, which is a
straightforward discretization of Eq.~\eqref{eq:topocharge_continuum} with
definite parity~\cite{DiVecchia:1981aev, DiVecchia:1981hh}:
\beq\label{eq:clover_charge}
Q_{\clov} = \frac{-1}{2^9 \pi^2}\sum_{x}\sum_{\mu\nu\rho\sigma=\pm1}^{\pm4}\varepsilon_{\mu\nu\rho\sigma}
\Tr\left\{\Pi_{\mu\nu}^{(1\times1)}(x)\Pi_{\rho\sigma}^{(1\times1)}(x)\right\}\ ,
\eeq
where $\Pi_{\mu\nu}^{(1\times1)}(x)$ is the plaquette and the Levi-Civita symbol with negative entries is defined by
$\varepsilon_{\mu\nu\rho\sigma}=-\varepsilon_{(-\mu)\nu\rho\sigma}$ and complete antisymmetry.
When computing the susceptibility $\chi=\langle Q^2\rangle/V$
using $Q_{\clov}$, both multiplicative~\cite{Campostrini:1988cy} and
additive renormalizations appear\footnote{The same is true for all the ``non geometric'' gluonic definitions of the topological charge~\cite{Vicari:2008jw}.}:
\beq
\chi_{\gluo} = Z_Q^2 \frac{\braket{Q^2_{\clov}}}{V} + M_{\ensuremath{\mathrm{add}}}.
\eeq
Such renormalizations are due to ultraviolet (UV) fluctations at the scale of the lattice
spacing and must be properly subtracted in order to recover the proper
continuum scaling of $\chi_{\gluo}$. To avoid dealing with
these renormalization constants, smoothing algorithms are commonly employed to
dampen UV fluctuations while leaving the topological content of the gauge
fields unchanged. Several methods have been proposed, such as
cooling~\cite{Berg:1981nw,Iwasaki:1983bv,Itoh:1984pr,Teper:1985rb,Ilgenfritz:1985dz,Campostrini:1989dh,Alles:2000sc},
smearing~\cite{APE:1987ehd, Morningstar:2003gk} and gradient
flow~\cite{Luscher:2009eq, Luscher:2010iy}, all giving consistent results when
properly matched~\cite{Alles:2000sc, Bonati:2014tqa, Alexandrou:2015yba}. 

To compute the gluonic susceptibility, in this work we adopt the cooling method, which is
numerically very convenient. Since even after cooling the topological charge is
non-integer (although its typical values get closer and closer to integers as
the lattice spacing is reduced) we adopt the following prescription to assign
an integer topological charge to
configurations~\cite{DelDebbio:2002xa,Bonati:2015sqt}:
\beq\label{eq:qgluo_alpharounded}
Q_{\gluo} = \round\left\{\alpha Q_{\clov}^{(\cool)}\right\}\ ,
\eeq
where ``$\round$'' means that the quantity $\alpha Q_{\clov}^{(\cool)}$ is rounded to the closest integer and where the parameter $\alpha$ is defined by
\beq
\alpha=\min_{x\ge 1} \left\langle\left[x Q_{\clov}^{(\cool)} - \round\left\{x Q_{\clov}^{(\cool)}\right\}\right]^2\right\rangle\ .
\eeq
The parameter $\alpha$ is thus chosen in order to center the peaks of the
distribution of $\alpha Q_{\clov}^{(\cool)}$ at integer
values, and the constraint $x\ge 1$ is required to exclude the trivial minimum
$x=0$. In the end, our gluonic susceptibility is:
\beq
\chi_{\gluo} = \frac{\braket{Q^2_{\gluo}}}{V}\ , \qquad V=a^4 N_t N_s^3\ .
\eeq
In our simulations we observe that after $n_{\cool}\sim 100$ cooling
steps $\chi_{\gluo}$ has reached a plateau for all explored lattice
spacings. For this reason, we compute the gluonic susceptibility for that
number of cooling steps in all cases. Slightly changing the value of $n_{\cool}$ resulted in no change in the obtained results for $\chi_{\gluo}$ for each lattice spacing.

It is also possible to adopt fermionic definitions of the
topological charge, based on the properties of the spectrum of the lattice
Dirac operator. In the continuum limit, according to the index theorem, it
would be sufficient to compute the sum of the chiralities of the zero-modes
$u_0$, as
\beq
Q = \Tr\{\gamma_5\} = \sum_{\text{zero modes}} u_0^\dagger \gamma_5 u_0 = n_+ - n_- .
\eeq
On the lattice, since no exact zero-mode exists when using a non-chiral
fermion discretization, every mode contributes and the trace becomes a sum
over all modes. To properly define the topological charge, we
introduce the projector on the eigenspace spanned by the eigenstates of
$iD_{\stag}[U^{(2)}]$ (i.e., the same operator we have included in our
lattice action) with eigenvalues $\vert \lambda \vert\leq M$:
\beq\label{eq:projectors}
\mathbb{P}_M \equiv \sum_{\vert \lambda \vert \le M} u_{\lambda} u_{\lambda}^{\dagger}\ , \qquad \qquad 
i D_{\stag}[U^{(2)}] u_{\lambda} = \lambda u_{\lambda}, \quad \lambda \in \mathbb{R}\ .
\eeq
Our SP definition of the bare topological charge is:
\beq\label{eq:SP_charge_bare}
Q_{\SP,\mathrm{bare}}^{(\stag)} = \frac{1}{n_t} \Tr\left\{\Gamma_5 \mathbb{P}_M \right\} = \frac{1}{n_t}\sum_{\vert \lambda \vert \le M} u_\lambda^\dagger \Gamma_5 u_\lambda\ , 
\qquad \qquad \Gamma_5 = \gamma_5^{(\stag)}\ ,
\eeq
where the factor $n_t = 2^{d/2}=2^2$ takes into account the taste degeneration of the
staggered spectrum.

As discussed in Ref.~\cite{Bonanno:2019xhg}, this definition is affected only
by a multiplicative renormalization, since the fast decay of the spectral
projector at infinity eliminates any additive renormalization. This multiplicative constant
can be expressed in terms of traces of $\Gamma_5$ and of the spectral projector
$\mathbb{P}_M$, thus, we are able to obtain a fully-spectral renormalized
definition of the topological charge~\cite{Bonanno:2019xhg}:
\beq\label{eq:SP_charge}
Q_{\SP}^{(\stag)} = Z_{\SP}^{(\stag)} Q_{\SP,\text{bare}}^{(\stag)},\\
\label{eq:SP_renorm_const}
Z_{\SP}^{(\stag)} = 
\sqrt{\frac{\braket{\Tr\left\{\mathbb{P}_M\right\}}}{\braket{\Tr\left\{\Gamma_5 \mathbb{P}_M \Gamma_5 \mathbb{P}_M\right\}}}}\ .
\eeq
The SP expression of the topological susceptibility can then be written as
\beq\label{eq:chi_SP_final}
\chi_{\SP}^{(\stag)} = {Z_{\SP}^{(\stag)}}^2 \frac{\left\langle {Q_{\SP,\text{bare}}^{(\stag)}}^2\right\rangle}{V} =  
\frac{1}{n_t^2}\frac{\braket{\Tr\{\mathbb{P}_M\}}}{\braket{\Tr\{\Gamma_5 \mathbb{P}_M \Gamma_5 \mathbb{P}_M \}}} \frac{\braket{\Tr\{\Gamma_5 \mathbb{P}_M\}^2}}{V}\ .
\eeq
Spectral traces can be computed by several means. For example, in
Refs.~\cite{Giusti:2008vb,Luscher:2010ik} noisy estimators are used to this
end. In this work we follow the same strategy of Ref.~\cite{Bonanno:2019xhg}
and we compute the first 200 smallest eigenvalues and eigenvectors of
$iD_{\stag}[U^{(2)}]$ using the \texttt{PARPACK} package~\cite{PARPACK}, so
that $\mathbb{P}_M$ is obtained directly from Eq.~\eqref{eq:projectors}. On our larger lattices, the
computational cost to obtain the 200 lowest-lying eigenvalues of
$iD_{\stag}[U^{(2)}]$ turned out to be about a factor of $\sim 20$ larger than the computational cost needed to perform a single RHMC step for the same lattice.

Once the eigenvalues and the eigenvectors of $iD_{\stag}[U^{(2)}]$ are obtained, spectral traces are then practically computed as follows:
\beq\label{eq:spectral_sums}
\begin{aligned}
\Tr\{\mathbb{P}_M\} &= \sum_{\vert\lambda\vert\le M} 1 = \nu(M),\\
\Tr\{\Gamma_5 \mathbb{P}_M\} &= \sum_{\vert\lambda\vert\le M} u_\lambda^\dagger \Gamma_5 u_\lambda,\\
\Tr\{\Gamma_5 \mathbb{P}_M \Gamma_5 \mathbb{P}_M\}&= \sum_{\vert\lambda\vert\le M}\sum_{\vert\lambda^\prime\vert\le M} \vert u_{\lambda}^\dagger \Gamma_5 u_{\lambda^\prime} \vert^2,
\end{aligned}
\eeq
where $\nu(M)$ is the total number of eigenmodes whose eigenvalues lie below $M$.

The cut-off mass $M$ is a free parameter, and from the index theorem we
know that its specific value is irrelevant in the continuum limit.  However,
$M$ should be kept constant in physical units as the continuum limit is
approached in order to observe the usual continuum scaling of the
susceptibility (i.e., $O(a^2)$ corrections in the staggered case), as discussed
in Ref.~\cite{Giusti:2008vb}. The cut-off mass renormalizes as a quark mass~\cite{Bonanno:2019xhg},
i.e., $M_R = Z_S^{-1} M$, where $Z_S$ is the renormalization constant of the
staggered flavor-singlet scalar fermionic density. For this reason, the ratio
between $M$ and any of the quark masses is a renormalization-group invariant
quantity: $M/m_f = M_R / m_f^{(R)}$. To keep $M_R$ constant in physical units,
it is thus sufficient to keep $\hat{M}/\hat{m}_f = M/m_f$ constant as we move
$\hat{m}_f$ along the LCP (where $\hat{M}\equiv aM$).  This strategy allows to completely avoid the
computation of $Z_S$ and will be the one adopted in this work.  When the
continuum limit is taken at fixed $M/m_f$, we thus expect the usual
scaling:
\beq\label{eq:SP_cont_scaling}
\chi_{\SP}^{(\stag)}(a,M/m_f) = \chi_{\SP} + c_{\SP}(M/m_f) a^2 + o(a^2)\ .
\eeq

Previous studies of QCD at zero temperature,
performed with twisted mass Wilson fermions and using twisted mass Wilson SP, have shown that
the SP determination of the topological susceptibility displays much smaller
lattice artifacts than the gluonic one~\cite{Alexandrou:2017bzk}. Although we
still do not have a complete quantitative understanding of this fact, an
intriguing possible interpretation exists (see also the discussion
at the end of Ref.~\cite{Bonanno:2019xhg}).  When using a non-chiral fermion
discretization we are sampling a lattice distribution that, for what concerns
the low-laying modes relevant for topology, can have quite large lattice
artifacts due to the explicit chiral symmetry breaking. If we use a gluonic
definition of the topological charge, we have lost any direct connection with
lattice chirality and the error introduced in the sampling propagates to the
measures.  If instead we use the SP definition built from the same Dirac
operator used to weight the configurations, there is the possibility that the
measure partially corrects the error in the sampling. Of course there is \emph{a
priori} no solid reason to exclude the possibility that the two different errors
sum up instead of canceling each other, so this argument can not be considered
in the present form conclusive. To understand to what extent such a
cancellation exists, it would be very interesting to perform a study using
different fermion discretizations in the generation of the configuration and in
the construction of the SP used to estimate the topological charge.

As a final remark we recall that, as already discussed in Ref.~\cite{Bonanno:2019xhg}, at finite temperature a possible ambiguity about the computation of the renormalization constant in Eq.~\eqref{eq:SP_renorm_const} could arise. In particular, one could wonder if $Z_\SP^{(\stag)}$ should be computed at finite $T$ or at zero $T$. In principle, such quantity should be computed in the latter case, which would require to perform, along each finite temperature simulation, a twin run with the same parameters but at zero temperature to compute $Z_\SP^{(\stag)}$. However, as already discussed in Ref.~\cite{Bonanno:2019xhg} in the quenched theory and as we will verify also in the following in the presence of dynamical fermions, the computation of $\chi_\SP^{(\stag)}$ at finite temperature gives fully consistent results both when $Z_\SP^{(\stag)}$ is obtained from the same finite-$T$ ensemble employed for the computation of $Q_{\SP}^{(\stag)}$ or from a corresponding ensemble at zero temperature, as long as the spectral traces appearing in Eq.~\eqref{eq:SP_renorm_const} are computed from the staggered spectrum obtained imposing periodic boundaries along the temporal direction for $D_\stag$\footnote{On the other hand, at zero temperature the choice of boundary conditions for the lattice Dirac operator along the temporal direction is expected to be irrelevant.}. Thus, in our finite temperature simulations, we will compute the spectrum of the Dirac operator both for periodic and anti-periodic boundary conditions along the temporal direction, and in Eq.~\eqref{eq:chi_SP_final} we will adopt the former to compute $Z_\SP^{(\stag)}$ and the latter to compute $Q_{\SP,\mathrm{bare}}^{(\stag)}$.

\subsection{Multicanonical algorithm}

The multicanonical approach consists in adding a topological bias potential
$V_{\topo}(Q_{\mc})$ to the action in order to enhance the
probability of visiting those topological sectors that would be otherwise
strongly suppressed:
\beq
S_{\YM}^{(L)} \rightarrow S_{\YM}^{(L)} + V_{\topo}(Q_{\mc})\ .
\eeq
The quantity $Q_{\mc}$ is a suitable discretization of the
topological charge, which in general differs from the one adopted for the
measurements $Q_\gluo$.  While the multicanonical algorithm is stochastically exact for
any choice of $Q_{\mc}$, to make it more efficient than the case
$V_{\topo}=0$ the discretization $Q_{\mc}$ has to satisfy a
couple of requirements. First, $Q_{\mc}$ must have a reasonable overlap with
the charge used in the measures, otherwise $V_{\topo}(Q_{\mc})$
would not work as a bias but just as noise. Second, it is also
important that $Q_{\mc}$ is not ``too peaked'' at integer values, to
avoid the need for very small integration steps in the Hybrid Monte Carlo.

The partition function in the presence of the topological potential $Z^{(\mc)}_{\LQCD}$ is the following:
\beq\label{eq:partition_function_bias}
Z^{(\mc)}_{\LQCD}=\int [dU] e^{-\left (S^{(L)}_\YM[U] + V_{\topo}(Q_{\mc})\right )} 
\det\left\{\mathcal{M}^{(\stag)}_{l}[U]\right\}^{\frac{1}{2}} \det\left\{\mathcal{M}^{(\stag)}_{s}[U]\right\}^{\frac{1}{4}}.
\eeq
For a generic observable $\mathcal{O}$, expectation values with respect to the original path-integral distribution $\braket{\mathcal{O}}$ can thus be exactly recovered through a simple reweighting procedure:
\beq\label{eq:reweighting_formula}
\braket{\mathcal{O}} = \frac{\braket{\mathcal{O}e^{V_{\topo}(Q_{\mc})}}_{\mc}}{\braket{e^{V_{\topo}(Q_{\mc})}}_{\mc}}\ ,
\eeq
where $\braket{\cdot}_{\mc}$ refers to expectation values computed in the presence of the topological bias.

\section{Results}\label{sec:results}

\subsection{Topological susceptibility at zero temperature}\label{sec:results_t_0}

At zero temperature, the value of the topological susceptibility in QCD can be reliably
computed using Chiral Perturbation Theory (ChPT)~\cite{DiVecchia:1980yfw,Leutwyler:1992yt,Mao:2009sy,Guo:2015oxa,GrillidiCortona:2015jxo,Luciano:2018pbj}, and the value obtained in this
way constitutes a useful benchmark for lattice determinations. Using the Next-to-Leading Order (NLO) expression
of Ref.~\cite{GrillidiCortona:2015jxo} one gets for the case $m_u = m_d$ at the physical point the estimate~\cite{Bonati:2015vqz}:
\beq\label{eq:chi_T0_ChPT}
\chi^{1/4}_{\text{ChPT}}=77.8(4)\text{ MeV}\ , \qquad (m_u/m_d=1)\ .
\eeq

Moreover, several gluonic determinations from the lattice have been reported in
the literature, see, e.g., Refs.~\cite{Bonati:2015vqz, Borsanyi:2016ksw}. We will use the computation of the zero temperature topological susceptibility as a test to validate our implementation, and to verify that the spectral projector approach has significantly smaller lattice artifacts than the standard gluonic one.

\begin{figure}[!htb]
\centering
\includegraphics[scale=0.5]{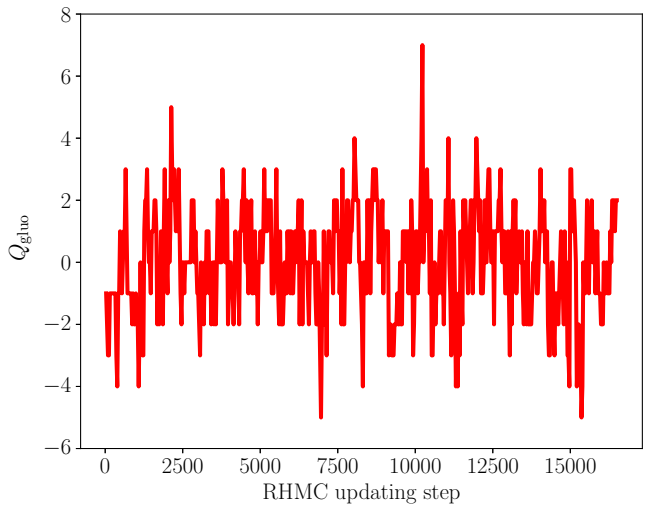}
\caption{Monte Carlo history of $Q_{\gluo}$, defined in
Eq.~\eqref{eq:qgluo_alpharounded}, for our run with $\beta=4.140$ at
$T\simeq0$, corresponding to the finest lattice spacing explored at this
temperature.}
\label{fig:MC_story_qgluo_T_zero}
\end{figure}

Our zero-temperature simulations have been performed on hypercubic $N_s^4$
lattices and the simulation parameters adopted are reported in
Tab.~\ref{tab:simulation_parameters_zero_t}. The lattice size was chosen so
that $L_s \equiv a N_s \sim 2.6$--$3.2$~fm, which is sufficient to keep
finite-size effects within our typical statistical errors~\cite{Bonati:2015vqz}.
At zero temperature, there is no need to introduce a bias potential in the
action, since several topological sectors are naturally explored during
the Monte Carlo evolution. As an example, we show the history of the topological charge $Q_{\gluo}$ for our finest lattice spacing in Fig.~\ref{fig:MC_story_qgluo_T_zero}.

\begin{table}[!htb]
\begin{center}
\begin{tabular}{|c|c|c|c|}
\hline
$\beta$ & $a$~[fm] & $\hat{m}_s \cdot 10^{2}$ & $N_s$ \\
\hline
3.750 & 0.1249 & 5.03 & 24 \\
3.850 & 0.0989 & 3.94 & 32 \\
3.938 & 0.0824 & 3.30 & 32 \\
4.020 & 0.0707 & 2.81 & 40 \\
4.140 & 0.0572 & 2.24 & 48 \\
\hline
\end{tabular}
\end{center}
\caption{Summary of simulation parameters for the $T\simeq 0$ runs, performed on
hypercubic $N_s^4$ lattices. The bare parameters $\beta$ and $\hat{m}_s$ and
the lattice spacings have been fixed according to the LCP determined in Refs.~\cite{Aoki:2009sc, Borsanyi:2010cj, Borsanyi:2013bia}, and
$\hat{m}_{l}$ is fixed through $\hat{m}_s/\hat{m}_{l}=m_s/m_l=28.15$.}
\label{tab:simulation_parameters_zero_t}
\end{table}

To compute $\chi$ using spectral projectors we have to choose the
cut-off mass $M$, which in the following computations will be normalized to the strange quark mass $m_s$. Since we know that in the continuum limit
only zero modes provide a non-vanishing contribution to the topological charge,
\emph{a priori} the optimal possibility would be to fix a value of $M/m_s$ large
enough to include in the spectral sums in Eqs.~\eqref{eq:spectral_sums} all the Would-Be
Zero Modes (WBZMs) of all the configurations, but small enough to
leave out all the other modes that become irrelevant in the continuum. In general, whether
this ``golden choice'' of $M/m_s$ exists or not depends on the amount
of explicit chiral symmetry breaking of the fermion discretization. However, in the
$T=0$ case we can already guess that such a sharp separation can not be present
in the spectrum. Indeed, due to the Banks-Casher relation, the spontaneous
breaking of chiral symmetry is associated to a proliferation of near-zero
modes.

\begin{figure}[!htb]
\centering
\includegraphics[scale=0.42]{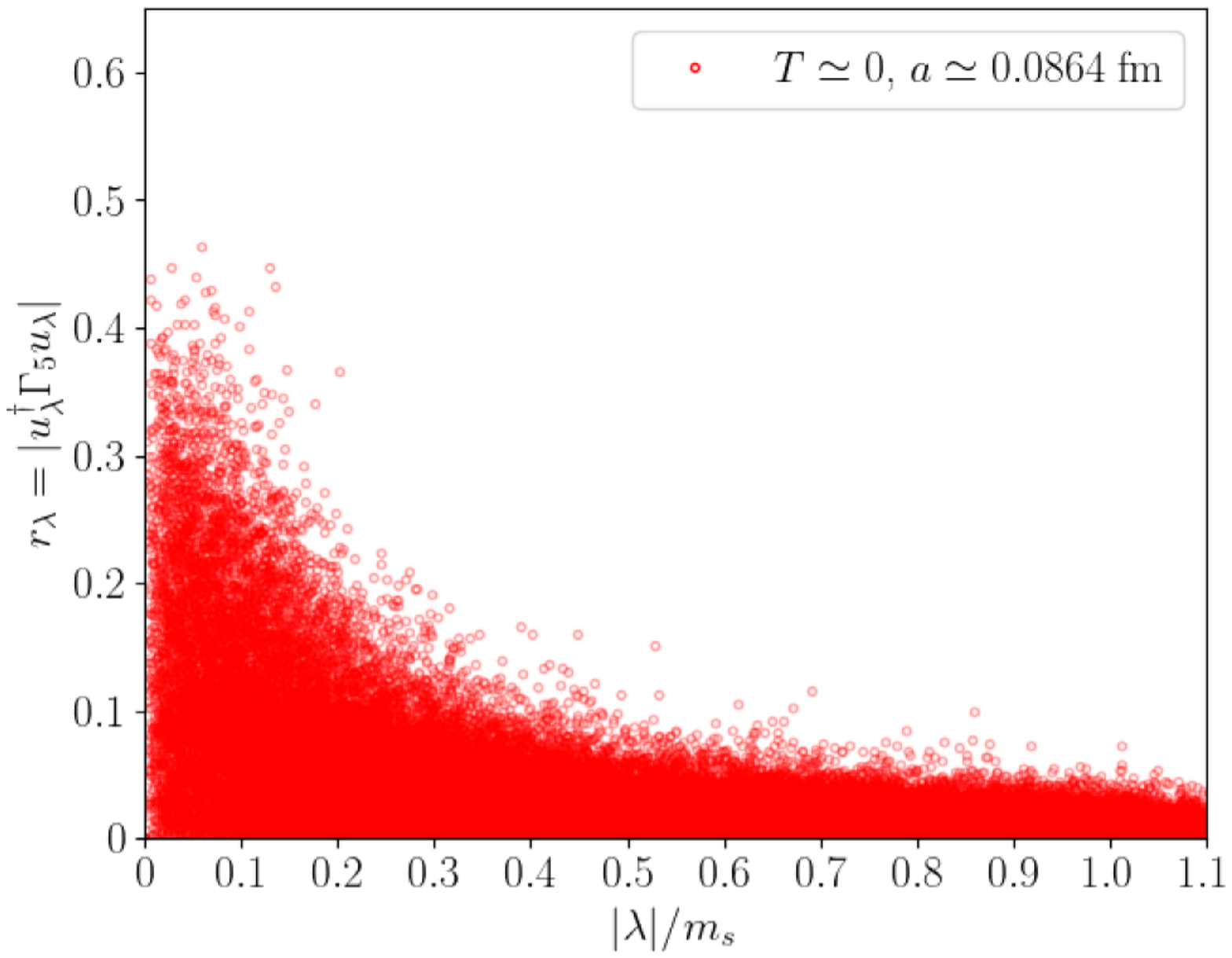}
\includegraphics[scale=0.42]{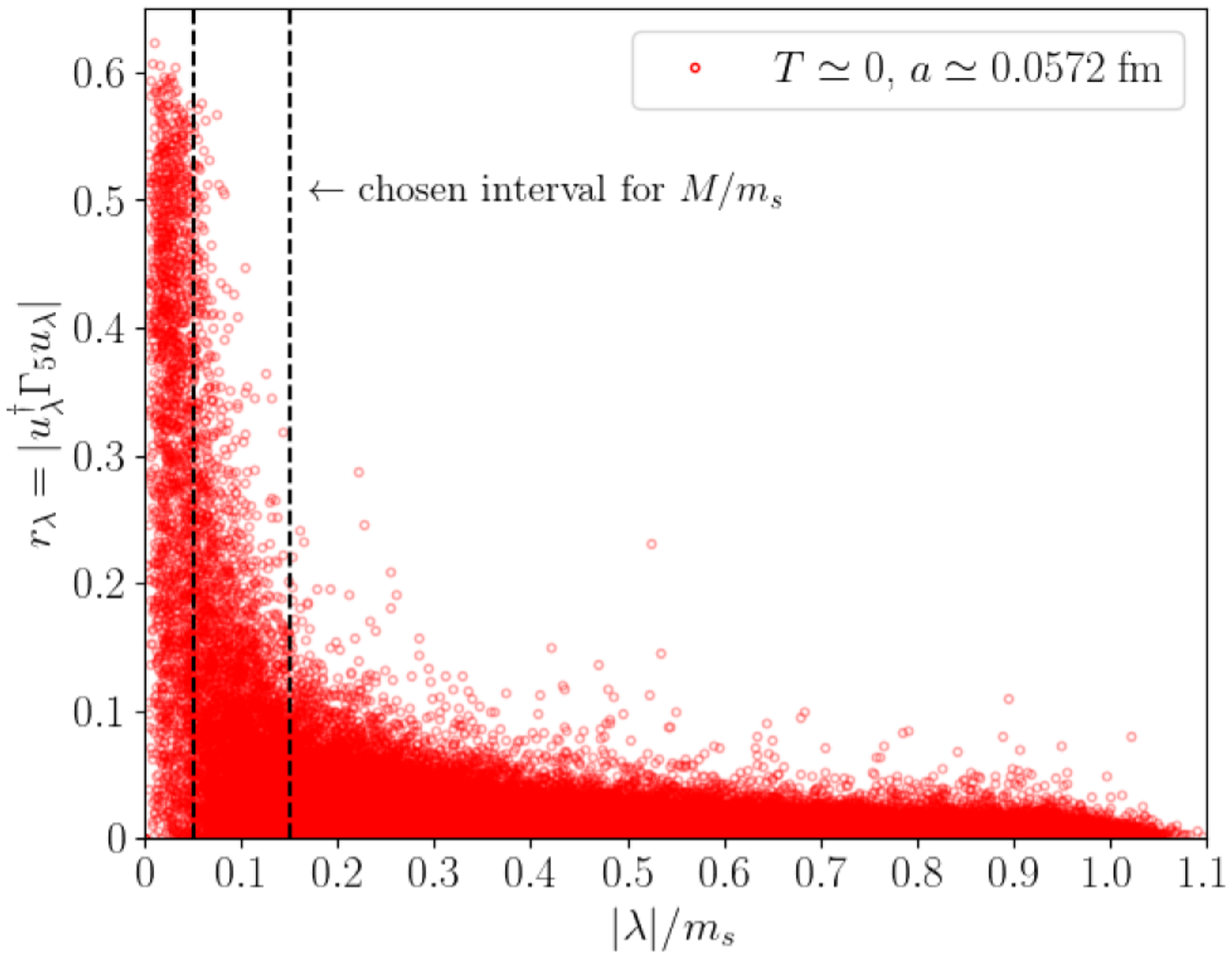}
\caption{Left: Scatter plot of chirality $r_\lambda$ vs $\vert\lambda\vert/m_s$ for a $32^4$ lattice with lattice spacing $a=0.0824$~fm.	Only the first $200$ eigenvalues (with the lowest magnitude) of $670$ configurations (taken every $30$ RHMC steps) are shown. Right: Scatter plot of chirality $r_\lambda$ vs $\vert\lambda\vert/m_s$ for a $48^4$ lattice with lattice spacing $a=0.0572$~fm. The two dashed vertical lines are set at $0.05$ and $0.15$ and denote our choice for the $M$-range. Only the first $200$ eigenvalues (with the lowest magnitude) of $540$ configurations (taken every $30$ RHMC steps) are shown.}
\label{fig:scatter_plot_zero_t}
\end{figure}

Using for the eigenmodes and the eigenvalues of $iD_{\stag}[U^{(2)}]$ the
notation $u_{\lambda}$ and $\lambda$ respectively (cf.~Eq.~\eqref{eq:projectors}), we can
associate to each mode its (absolute) chirality $r_\lambda \equiv \vert u^\dagger_\lambda
\Gamma_5 u_\lambda \vert$. In the continuum $r_{\lambda}=1$ if $\lambda=0$ and
$r_{\lambda}=0$ if $\lambda\neq 0$, while on the lattice we generically have
$0<r_{\lambda}<1$. However, $r_{\lambda}$ can be used as a figure of merit to
identify WBZMs. In Fig.~\ref{fig:scatter_plot_zero_t} on the right we report a scatter plot
of $r_{\lambda}$ against $\vert\lambda\vert/m_{s}$ for the first 200 low-lying modes of
540 $T\approx 0$ configurations (taken every 30 RHMC steps) generated for our finest lattice spacing at this temperature, $a\simeq0.0572$~fm. It is evident that a sharp separation between WBZMs and non-chiral modes does not exist.

Having seen that the golden strategy is unfeasible, we identified at
the finest lattice spacing a range of cut-offs $M/m_s$ chosen to reasonably
include in the spectral sums all the modes $\vert\lambda\vert\le M$ that look ``chiral enough''; in the following we will simply refer to this range as the ``$M$-range''. Continuum extrapolation is then performed for several values of $M/m_s$ chosen within this range, and the residual variability will be taken as a systematic of the extrapolation procedure.

The identification of the $M$-range has been done for the finest lattice spacing because, being the closest point to the continuum limit, the distinction between chiral and non-chiral mode is clearer. For comparison, in Fig.~\ref{fig:scatter_plot_zero_t} on the left we also report the scatter plot of $r_\lambda$ against $\vert\lambda\vert/m_s$ for our intermediate lattice spacing at this temperature, $a\simeq 0.0824$~fm. While in this case the identification of a reasonable range for the cut-off appears difficult, for the finest lattice spacing it is clear that for $\vert\lambda\vert/m_s\approx 0.1$ there is a change of regime. Therefore, we chose as the $M$-range the interval $[0.05, 0.15]$.

\begin{figure}[!htb]
\centering
\includegraphics[scale=0.5]{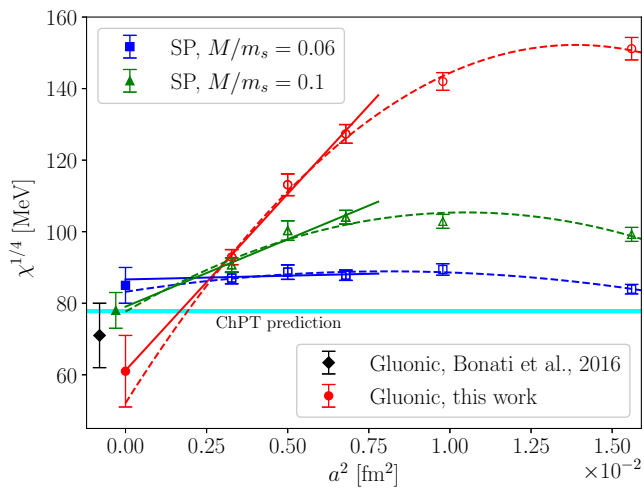}
\caption{Extrapolation towards the continuum limit of $\chi^{1/4}$ for $T\approx 0$. The
horizontal band shows NLO ChPT prediction in Eq.~\eqref{eq:chi_T0_ChPT}. Our gluonic determinations have been determined through the alpha-rounded clover definition computed on cooled configurations after $n_\cool=100$ cooling steps. The diamond full point represents the gluonic determination reported in Ref.~\cite{Bonati:2015vqz}.}
\label{fig:continuum_limit_zero_t}
\end{figure}

Once the $M$-range has been determined, it is possible to
extract the continuum limit at fixed value of $M/m_s$. In
Fig.~\ref{fig:continuum_limit_zero_t}, we compare the continuum limit obtained
using the standard gluonic definition with those obtained through SP for two
different values of $M/m_s$. In all cases continuum extrapolation is
performed in two different ways: a linear fit in $a^2$ restricted to the three
smallest lattice spacings, and a quadratic fit in $a^2$ in the whole range, cf.~Eq.~\eqref{eq:SP_cont_scaling}. Results obtained varying the fit range and the fit function appear to be in very good agreement in all cases, but we observe that SP continuum extrapolations are less sensitive to the fitting procedure adopted, cf.~Fig.~\ref{fig:continuum_limit_zero_t}.

As a matter of fact, if we compare the magnitude of the $O(a^2)$ lattice artifacts affecting the two discretizations, we observe that SP estimates have a much faster convergence
towards the continuum limit. If we denote by
$c_{\SP}$ the coefficient of the $O(a^2)$ correction to $\chi$ for the SP
approach (see Eq.~\ref{eq:SP_cont_scaling}) and by $c_{\gluo}$ the
corresponding coefficient for the gluonic definition, we have
$c_{\SP}(0.06)/c_{\gluo}\sim10^{-2}$ and
$c_{\SP}(0.1)/c_{\gluo}\sim0.3$ for the two particular values
of $M/m_s$ shown in Fig.~\ref{fig:continuum_limit_zero_t}.

\begin{figure}[!htb]
\centering
\includegraphics[scale=0.42]{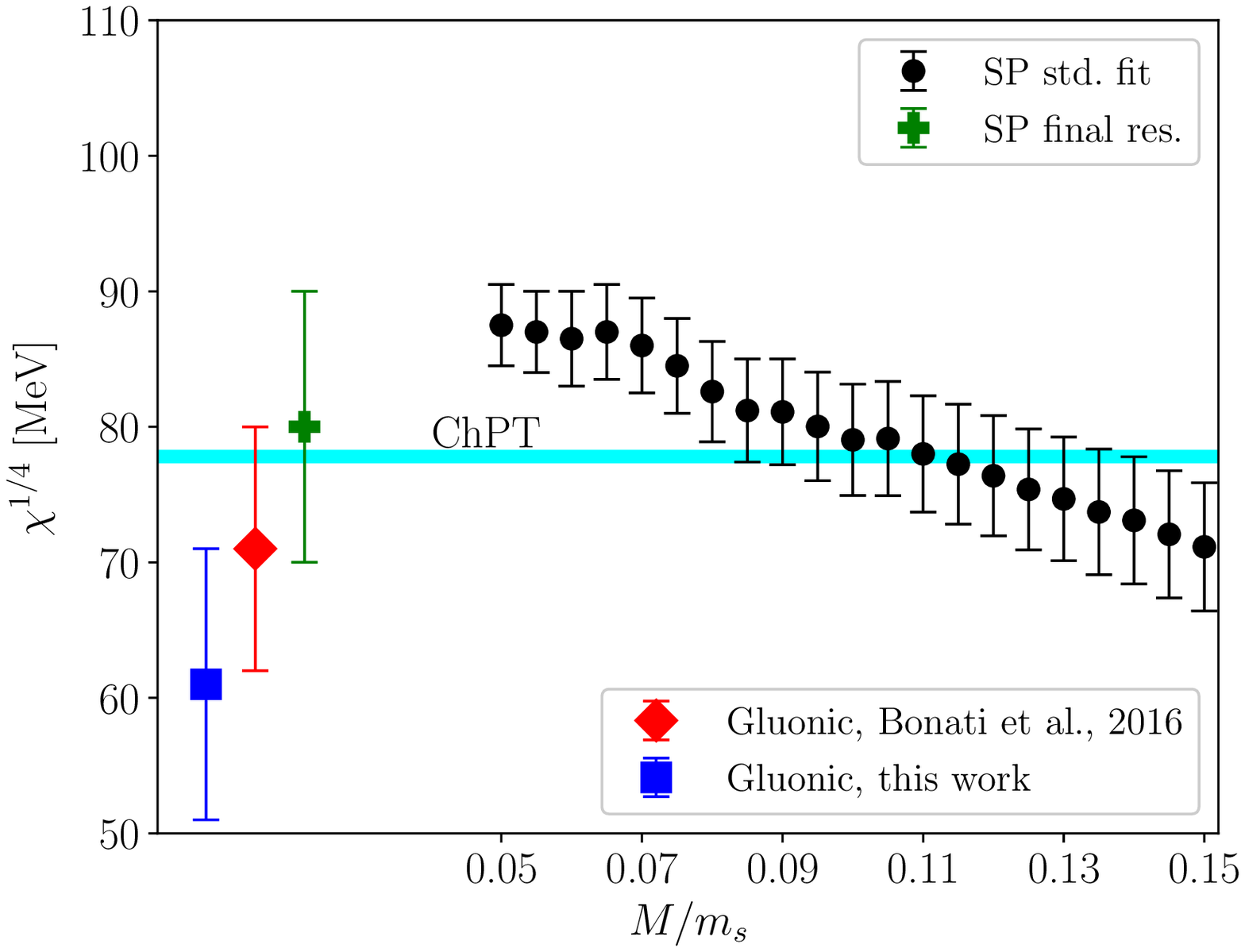}
\hspace{2mm}
\includegraphics[scale=0.42]{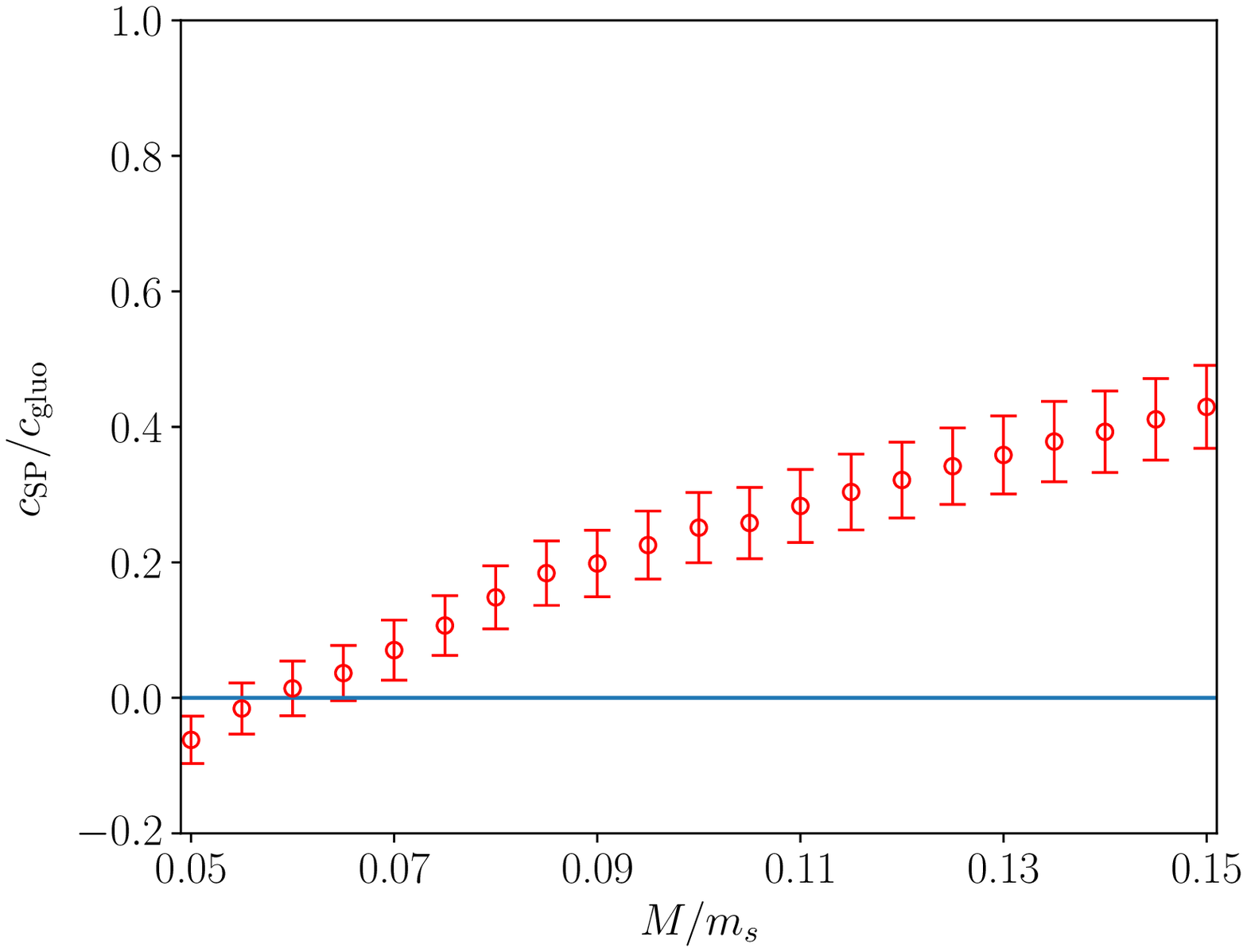}
\caption{Left: continuum limits of $\chi^{1/4}$ obtained at $T\approx 0$ from spectral projectors for several values of the cut-off $M/m_s$ chosen within the $M$-range. Each error bar refers to the continuum extrapolation obtained fitting the three finest lattice spacings with the fit function reported in Eq.~\eqref{eq:SP_cont_scaling}. The cross point represents our final SP determination of $\chi^{1/4}$, which includes any residual systematic related to the choice of $M/m_s$. The diamond full point represents the gluonic determination reported in Ref.~\cite{Bonati:2015vqz}. The horizontal band displays the NLO ChPT result of Eq.~\eqref{eq:chi_T0_ChPT}. Right: behavior of $c_{\SP}/c_{\gluo}$ as a function of the cut-off $M/m_s$ within the $M$-range. A straight horizontal line is set at $0$.}
\label{fig:systematic_chi_zero_t}
\end{figure}

In order to give a final result for $\chi^{1/4}$ from SP, it is necessary to
correctly assess the systematic error related to the choice of $M/m_s$. In
Fig.~\ref{fig:systematic_chi_zero_t} on the left we show how the SP continuum
extrapolation obtained from a linear fit in $a^2$ of the three finest lattice spacings depends on the choice of $M/m_s\in[0.05,0.15]$. These
determinations are all compatible among each other, however, we observe that,
as $M/m_s$ grows, the central value of the continuum extrapolation tends to
drift downward. For this reason, our final result $\chi^{1/4}_{\SP} = 80(10)$~MeV is obtained by choosing a confidence interval that keeps this systematic variation into account, cf.~Fig.~\ref{fig:systematic_chi_zero_t}. As for the gluonic result, instead, we estimated the final error from the systematic variation of the extrapolation when changing the fit function and the fit range. These final results are in good agreement among themselves and also with the gluonic determination reported in Ref.~\cite{Bonati:2015vqz} and with the ChPT prediction in Eq.~\eqref{eq:chi_T0_ChPT}, as shown in Fig.~\ref{fig:systematic_chi_zero_t}.

To test the solidity of our results, we performed a further consistency check of our continuum extrapolation procedure, consisting of a common continuum extrapolation of the gluonic results and the SP determinations at fixed $M$. In this case it is important to include $O(a^4)$ corrections to the gluonic determination and, since SP data show milder lattice artifacts with respect to the gluonic ones, the final result turns out to be practically indistinguishable from the SP continuum extrapolation at the same $M$ value. This test constitutes a non-trivial consistency check of the adopted procedure, and we report as our final estimate
\beq
\chi^{1/4}_{\SP}(T=0) = 80(10)~\text{MeV}.
\eeq
The same procedure to assess the final error on $\chi^{1/4}$ will also be applied at finite temperature.

Finally, we show in the right plot of Fig~\ref{fig:systematic_chi_zero_t} the
dependence of $c_{\SP}/c_{\gluo}$ on the cut-off $M/m_s$ within the $M$-range.
We observe that the SP discretization is affected by smaller
lattice artifacts compared to the gluonic definition, and that corrections to
the continuum limit grow as $M/m_s$ is increased, getting closer to those of
the gluonic discretization. This is due to the fact that, as $M/m_s$ grows, the
number of irrelevant non-chiral modes, which are
more affected by UV cut-off effects, included in the spectral sums grows too.

\subsection{Topological susceptibility at finite temperature}\label{sec:results_t_430}

We computed the topological susceptibility for five temperature values in the
high temperature phase of QCD, and specifically for $T\simeq 230~\text{MeV}$,
$T\simeq 300~\text{MeV}$, $T\simeq 365~\text{MeV}$, $T\simeq 430~\text{MeV}$
and $T\simeq 570~\text{MeV}$. In this section, we will discuss the details of
the $T=430~\text{MeV}$ case; a similar analysis has been carried out also for
the other temperature values, and the results obtained in these cases are
reported in Appendix~\ref{sec:appendix_finite_T_res}.

Simulations at $T\simeq 430$~MeV have been performed on $N_s^3\times N_t$
lattices following the same LCP already used in the zero temperature case; simulation parameters are
reported in Tab.~\ref{tab:simulation_parameters_T_430}. The spatial extent of
the lattice was chosen to ensure an aspect ratio not smaller than 3.
\begin{table}[!htb]
\begin{center}
\begin{tabular}{|c|c|c|c|c|}
\hline
$\beta$ & $a$~[fm] & $\hat{m}_s \cdot 10^{-2}$ & $N_s$ & $N_t$ \\
\hline
4.140 & 0.0572 & 2.24 & 32 & 8  \\
4.280 & 0.0458 & 1.81 & 32 & 10 \\
4.385 & 0.0381 & 1.53 & 36 & 12 \\
4.496 & 0.0327 & 1.29 & 48 & 14 \\
4.592 & 0.0286 & 1.09 & 48 & 16 \\
\hline
\end{tabular}
\end{center}
\caption{Simulation parameters for the runs at $T=1/\left(a
N_t\right)\simeq 430~\text{MeV}\simeq 2.8~T_c$. The bare parameters $\beta$,
$\hat{m}_s$ and the lattice spacings have been fixed according to the LCP determined in Refs.~\cite{Aoki:2009sc, Borsanyi:2010cj, Borsanyi:2013bia}, and
$\hat{m}_{l}$ is fixed through $\hat{m}_s/\hat{m}_{l}=m_s/m_l=28.15$.}
\label{tab:simulation_parameters_T_430}
\end{table}

Simulations at finite temperature, unlike those at $T=0$, have been performed
adopting the multicanonical algorithm. Our implementation of the multicanonical
algorithm closely follows the one already adopted in
Ref.~\cite{Bonati:2018blm}. The quantity $Q_{\mc}$ entering the
topological potential is the clover discretization~\eqref{eq:clover_charge} of
the topological charge computed after $n_{\mc}$ levels of stout smearing, which
allows to adopt the RHMC algorithm also in the presence of the topological
potential. The values of $n_{\mc}$ used ranged from $n_{\mc}=20$ for the coarsest lattice
spacing to $n_{\mc}=10$ for the finest one, while the isotropic smearing parameter was
always fixed to $\rho_{\mc}=0.1$. The functional form of the topological potential
adopted was~\cite{Bonati:2018blm}:
\beq\label{eq:topobias}
V_{\topo}(x)=
\begin{cases}
-\sqrt{ (Bx)^2 + C}, &\quad \vert x \vert < Q_{\max}\ ,\\
-\sqrt{ (BQ_{\max})^2 + C}, &\quad \vert x \vert \geq Q_{\max}\ ,
\end{cases}
\eeq
and in our simulations, $Q_{\max}=3$ turned out to be sufficient to
observe a dramatic enhancement in the number of fluctuations of
$Q_{\gluo}$. The free parameters $B$ and $C$ have instead been tuned
through short preliminary runs by requiring the Monte Carlo histories of the measured
topological charge $Q_{\gluo}$ to
uniformly explore the interval $[-Q_{\max},\,Q_{\max}]$.

\begin{figure}[!htb]
\centering
\includegraphics[scale=0.42]{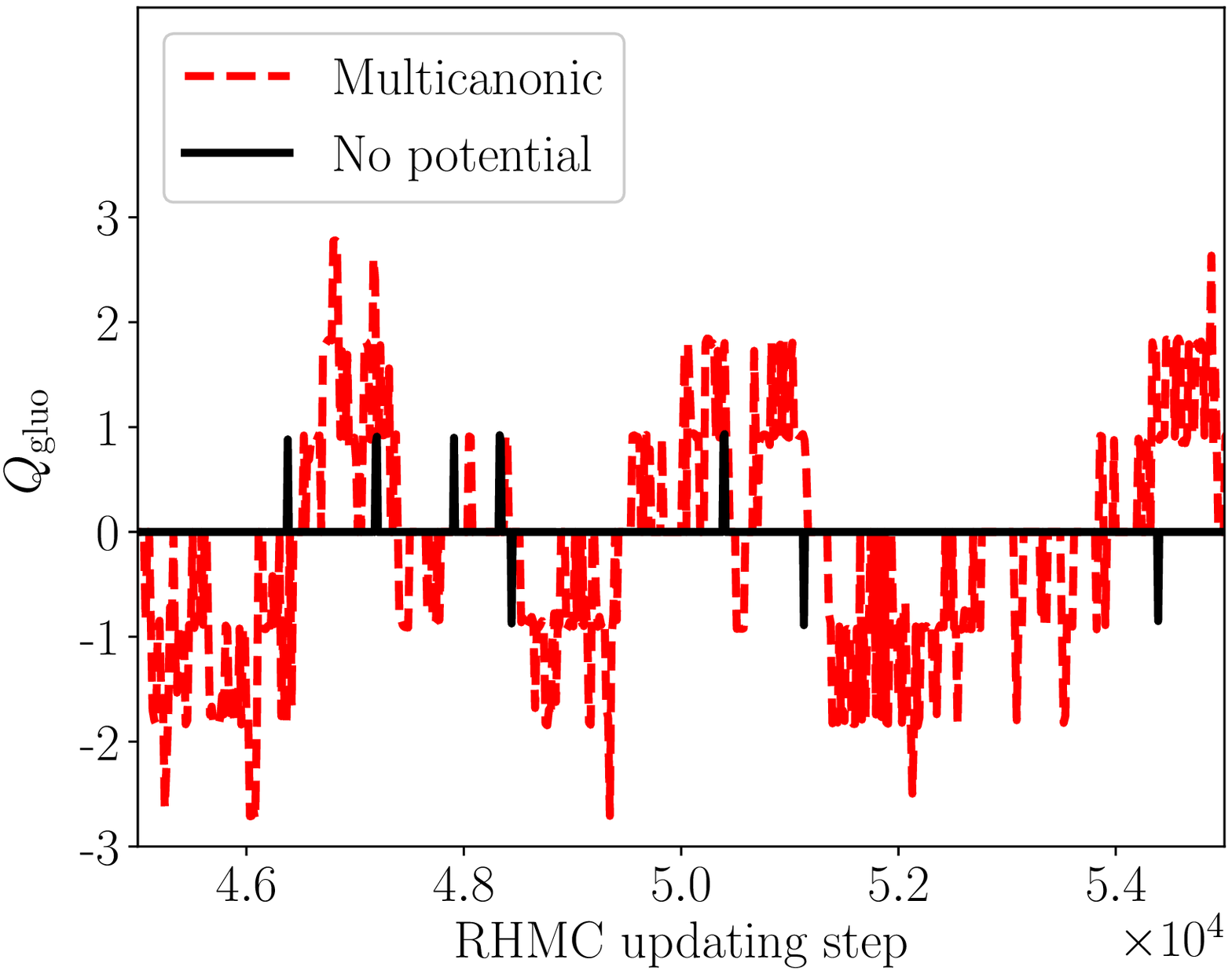}
\hspace{2mm}
\includegraphics[scale=0.42]{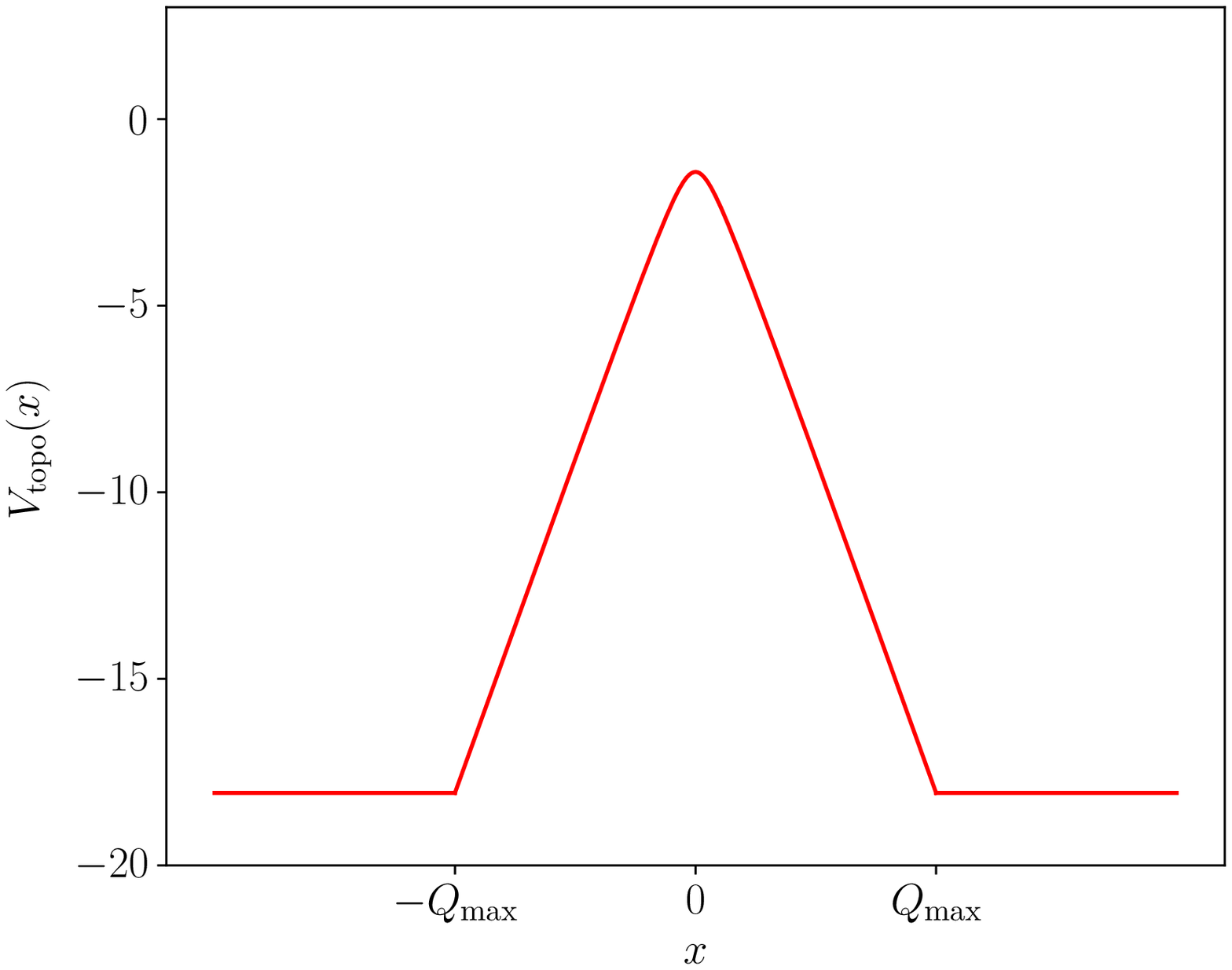}
\caption{Left: Comparison between the Monte Carlo histories of
$Q_{\gluo}$ obtained with and without bias potential for our run with
$\beta=4.140$ at $T\simeq430~\text{MeV}$. Since for this point one RHMC step in the presence
of the potential requires a $\sim 60\%$ larger numerical effort, to make the
comparison fair we expressed the Monte Carlo time on the horizontal axis in
units of one standard RHMC step in both cases. Right: Functional form~\eqref{eq:topobias} of the bias potential. In this case $B=6$, $C=2$ and
$Q_{\max}=3$.}
\label{fig:multicanonical_algorithm}
\end{figure}

The improvement obtained with the multicanonic algorithm is exemplified in
Fig.~\ref{fig:multicanonical_algorithm}. Without any bias potential,
$Q_{\gluo}$ assumes a non-zero value only a handful of times, while
much more fluctuations are observed as the topological potential is switched
on\footnote{In Ref.~\cite{Bonati:2018blm} it has been checked, whenever a simulation without multicanonic algorithm was feasible too, that the adoption of the multicanonic algorithm does not introduce any bias and yields the same result for topological observables.}. This allows to dramatically improve the accuracy with which $\chi$ can be
computed with a given machine time budget.

\begin{figure}[!htb]
\centering
\includegraphics[scale=0.42]{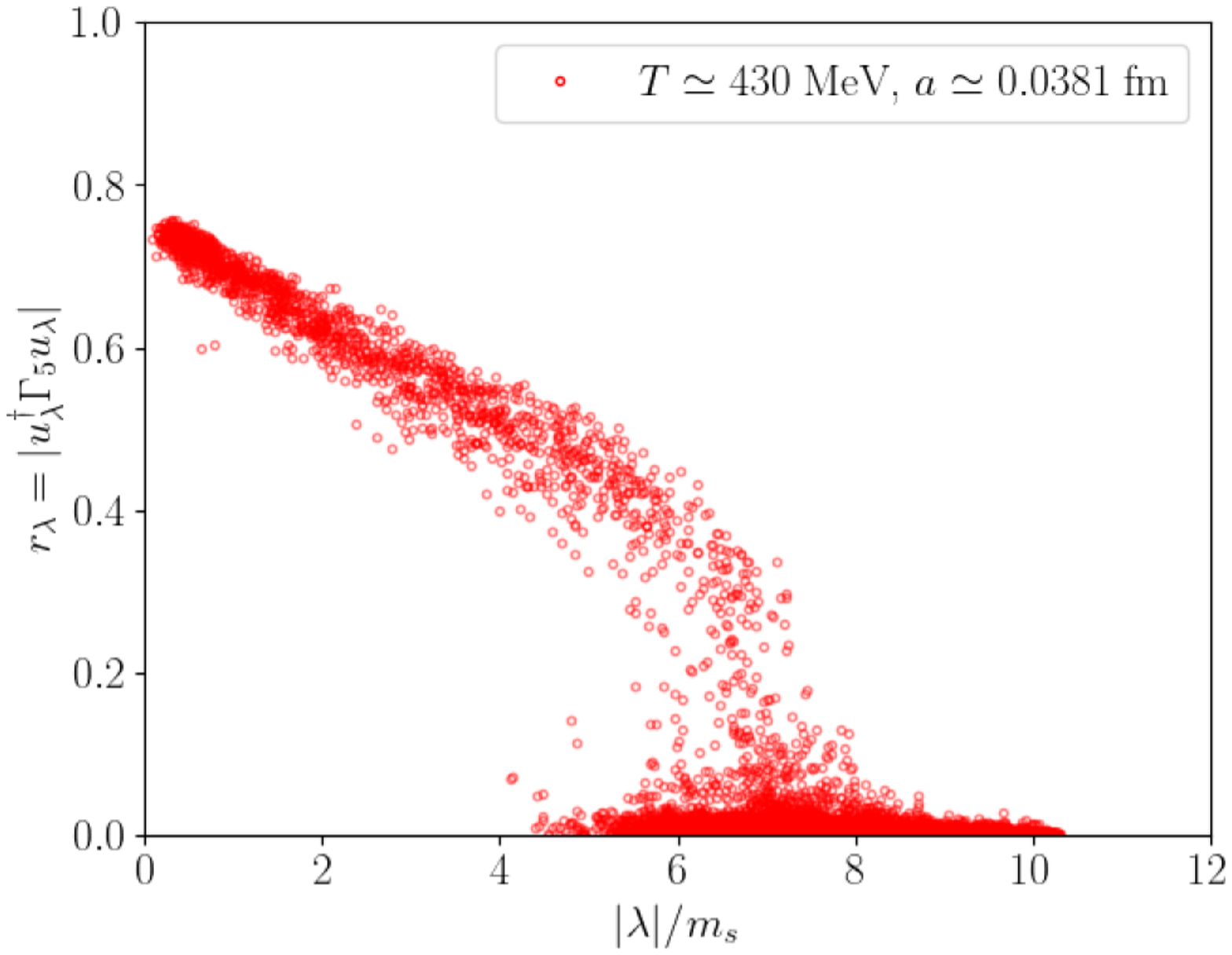}
\includegraphics[scale=0.42]{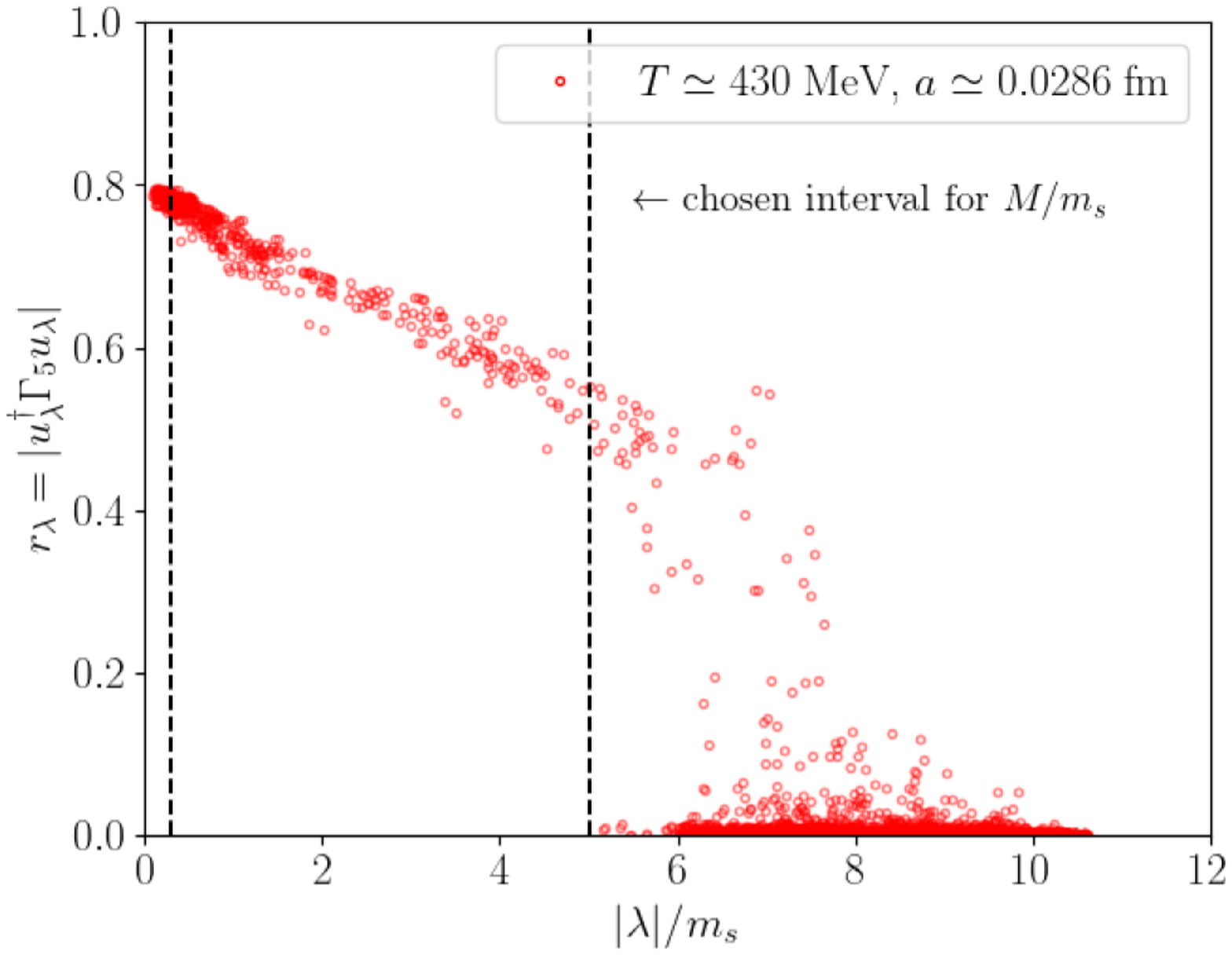}
\caption{Left: scatter plot of the chirality $r_\lambda$ vs $\vert\lambda\vert/m_s$ for our run with $a\simeq0.0381$~fm at $T\simeq430$~MeV. Only the first $200$ eigenvalues (with the lowest magnitude) of $227$ configurations (taken every $30$ RHMC steps) are shown. Right: scatter plot of the chirality $r_\lambda$ vs $\vert\lambda\vert/m_s$ for our run with $a\simeq0.0286$~fm at $T\simeq430$~MeV. The two dashed vertical lines are set at $0.3$ and $5$ and delimit the chosen $M$-range. Only the first $200$ eigenvalues (with the lowest magnitude) of $184$ configurations (taken every $1200$ RHMC steps) are shown.}
\label{fig:scatter_plot_t_430}
\end{figure}

In order to compute the topological susceptibility, we follow the same procedure already discussed for the $T\simeq 0$ case in
Sec.~\ref{sec:results_t_0}. The first step consists of determining a reasonable interval for $M/m_s$ by studying the scatter plot of the chiralities for the finest lattice spacing available, which is shown in Fig.~\ref{fig:scatter_plot_t_430} on the right.

With respect to the $T\simeq 0$
case, at finite temperature the separation between high and low-chirality modes
is more evident, as two almost disconnected clusters can be observed
with $r_\lambda\sim 0.7-0.8$ and $r_\lambda \lesssim 0.5$. However, also in
this case, the isolation of WBZMs is an ambiguous task, as we also observe a
sparsely-populated group of modes in between, with $0.5\lesssim
r_\lambda\lesssim0.7$.

\begin{figure}[!htb]
\centering
\includegraphics[scale=0.5]{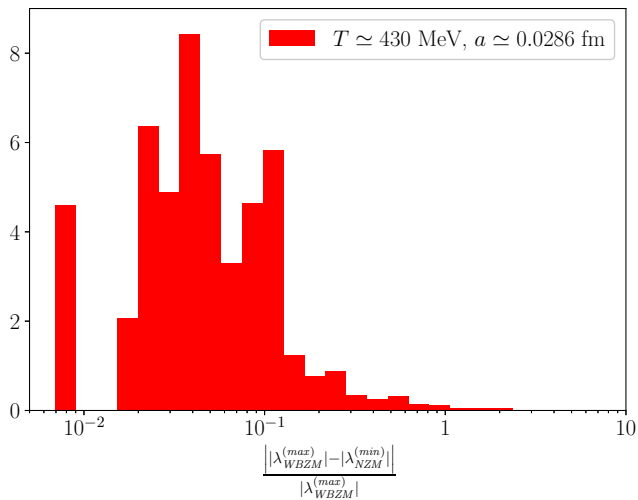}
\caption{Histogram of the absolute value of the relative separation between the magnitude of the largest eigenvalue of the candidate WBZMs $\vert\lambda_{\ensuremath{\mathrm{WBZM}}}^{(\max)}\vert$ and the smallest eigenvalue of the candidate NZMs $\vert\lambda_{\ensuremath{\mathrm{NZM}}}^{(\min)}\vert$. The identification of WBZMs was tentatively obtained, configuration by configuration, considering the first smallest $n_t \vert Q_{\gluo}\vert$ eigenvalues of $iD_{\stag}$. The plot refers to the finest lattice spacing $a\simeq0.0286$~fm explored at $T\simeq430$~MeV.}
\label{fig:wbzm_separation_histo}
\end{figure}

Such an ambiguity is not a characteristic of the whole configuration ensemble, but already arises configuration by configuration. In Fig.~\ref{fig:wbzm_separation_histo}, we plot the histogram of the separation between candidate WBZMs and Non-Zero Modes (NZMs) for the same sample represented in the scatter plot of Fig.~\ref{fig:scatter_plot_t_430} on the right.

The candidate WBZMs were identified from the value of the gluonic topological charge: more precisely we assumed the first $n_t\vert Q_\gluo\vert$ lowest-lying eigenmodes of $iD_{\stag}$ to be the WBZMs (the factor $n_t=4$ takes into account the four tastes of staggered fermions). This strategy is the same adopted in Ref.~\cite{Borsanyi:2016ksw} to identify WBZMs for their reweighting procedure, and it is justified on the basis of the index theorem for staggered fermions in the continuum $n_t Q = n_+ - n_-$, supplemented by the assumption that $n_-=0$ ($n_+=0$) if $Q>0$ ($Q<0$), which should be approximately true if the lattice volume is not too large. However, we observe that the typical relative separation between WBZMs and NZMs is of the order of $10^{-2}\div 10^{-1}$, cf.~Fig.~\ref{fig:wbzm_separation_histo}, meaning that a sharp separation between WBZMs and NZMs cannot be unambiguously established already at the level of the single configuration.

Thus, we cautiously choose our
cut-off masses in the interval $M/m_s\in[0.3, 5]$ in order to include in our spectral sums all modes with $\vert\lambda\vert\le M$ and $r_\lambda \gtrsim 0.5$. Again, the $M$-range has been identified for the finest lattice spacing available at this temperature, $a\simeq0.0286$~fm. However, we observe that at finite temperature also intermediate lattice spacings would lead approximately to the same choice. For comparison, in Fig.~\ref{fig:scatter_plot_t_430} on the left, we also report the scatter plot of $r_\lambda$ against $\vert\lambda\vert/m_s$ for an intermediate lattice spacing at this temperature, $a\simeq0.0381$~fm.

\begin{figure}[!htb]
\centering
\includegraphics[scale=0.5]{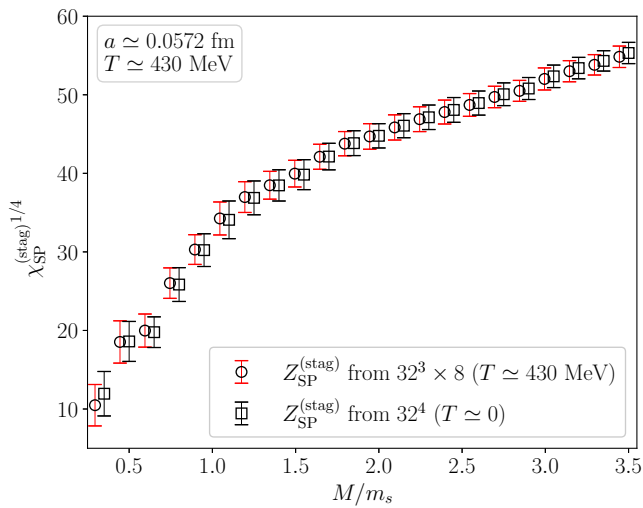}
\caption{Comparison of ${\chi_\SP^{(\stag)}}^{1/4}$ as a function of $M/m_s$ for $a\simeq 0.0572$~fm and at $T\simeq 430$~MeV when the multiplicative renormomalization constant $Z_{\SP}^{(\stag)}$ is computed on our finite temperature ensemble (on a $32^3\times 8$ lattice) or on a corresponding zero-temperature ensemble (on a $32^4$ lattice). The quantity $Z_{\SP}^{(\stag)}$ was computed in both cases using the staggered spectrum obtained imposing periodic boundaries along the temporal direction for $D_\stag$.}.
\label{fig:renorm_const_comp}
\end{figure}

In Fig.~\ref{fig:renorm_const_comp} we compare results for $\chi_\SP^{1/4}$ at $a\simeq 0.0572$~fm obtained, respectively, by computing the multiplicative renormalization constant $Z_{\SP}^{(\stag)}$ in Eq.~\eqref{eq:SP_renorm_const} from our finite temperature ensemble generated on a $32^3\times8$ lattice and from a corresponding zero temperature ensemble on a $32^4$ lattice. We recall that, while the computation of the spectral traces appearing in the definition of $Q_{\SP,\mathrm{bare}}^{(\stag)}$ in Eq.~\eqref{eq:SP_charge_bare} has been done from the staggered spectrum of $D_\stag$ where anti-periodic boundaries are imposed along the temporal direction, the computation of $Z_{\SP}^{(\stag)}$ in Eq.~\eqref{eq:SP_renorm_const} is done from the staggered spectrum in the presence of periodic boundaries. While this choice matters at finite temperature, it is irrelevant at zero temperature.

As Fig.~\ref{fig:renorm_const_comp} shows, both ways of computing $\chi_{\SP}^{1/4}$ give perfectly consistent results, in agreement with results obtained at finite temperature in the quenched theory~\cite{Bonanno:2019xhg}. For this reason, for all lattice spacings we computed both $Z_\SP^{(\stag)}$ and $Q_{\SP}^{(\stag)}$ from our finite temperature ensembles, where the staggered spectrum entering the spectral definition of these quantities was obtained, respectively, imposing periodic and anti-periodic boundaries along the time direction for $D_{\stag}$.

\begin{figure}[!htb]
\centering
\includegraphics[scale=0.5]{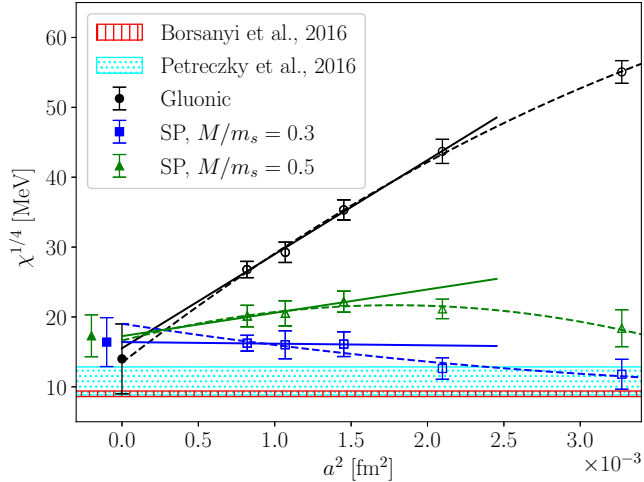}
\caption{Comparison of the continuum limits of $\chi^{1/4}$ for $T\simeq430$~MeV obtained with the gluonic and the SP discretizations. Vertically-hatched and dotted-hatched bands display the values of $\chi^{1/4}$ obtained for this temperature in, respectively, Refs.~\cite{Borsanyi:2016ksw, Petreczky:2016vrs}. The former was temperature-interpolated according to the DIGA prediction $\chi^{1/4} \sim T^{-2}$ and the isospin-breaking factor was removed. The latter was mass-extrapolated according to $\chi^{1/4}\sim m_\pi$.}
\label{fig:continuum_limit_t_430}
\end{figure}

In Fig.~\ref{fig:continuum_limit_t_430}, we compare the continuum limits of
$\chi^{1/4}$ obtained via the gluonic and the SP definitions, using the same
fit function in Eq.~\eqref{eq:SP_cont_scaling} already adopted for the $T=0$ case. For the SP case,
we show results for the values $M/m_s=0.3$ and $0.5$. As for the
zero-temperature case, also at finite temperature we observe that the SP definition allows to
achieve a sensible reduction of the magnitude of $O(a^2)$ corrections with
respect to the gluonic case, $c_{\SP}(0.3)/c_{\gluo}\sim 5
\times 10^{-2}$ and $c_{\SP}(0.5)/c_{\gluo}\sim 10^{-1}$,
allowing for a better control over systematics related to the continuum
extrapolation, as we observe a very good agreement in the obtained extrapolations when varying the fit range and the fit function. Moreover, we observe a good agreement among the gluonic and the SP determinations, cf.~Fig.~\ref{fig:continuum_limit_t_430}.

\begin{figure}[!htb]
\centering
\includegraphics[scale=0.42]{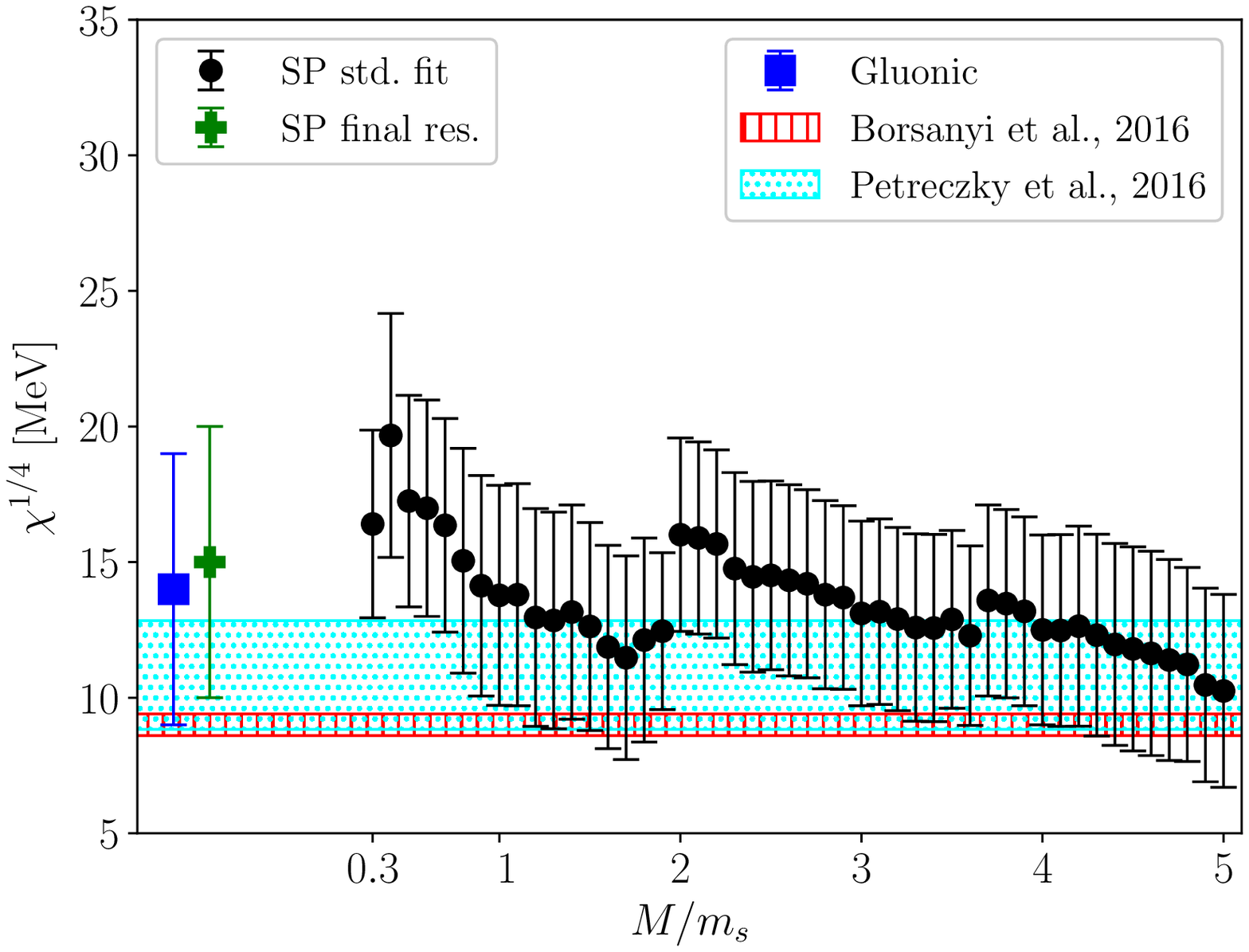}
\hspace{2mm}
\includegraphics[scale=0.42]{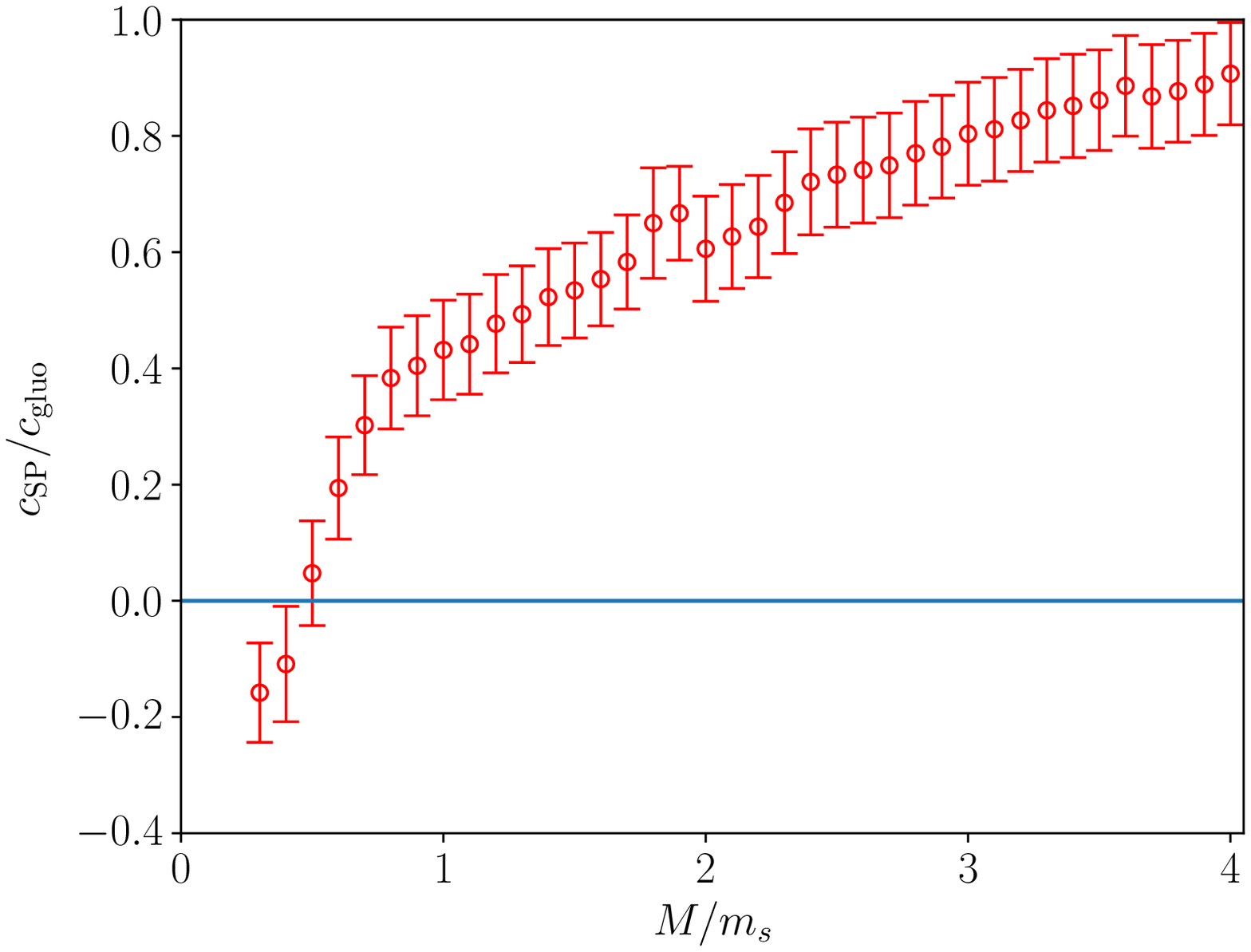}
\caption{Left: continuum limits of $\chi^{1/4}$ obtained at $T\simeq 430~\text{MeV}$ from spectral projectors for several values of the cut-off $M/m_s$ within the $M$-range. Each error bar refers to the continuum extrapolation obtained fitting the three finest lattice spacings with the fit function reported in Eq.~\eqref{eq:SP_cont_scaling}. The cross point represents our final SP determination of $\chi^{1/4}$, which includes any residual systematic related to the choice of $M/m_s$. Vertically-hatched and dotted-hatched bands display the values of $\chi^{1/4}$ obtained for this temperature in, respectively, Refs.~\cite{Borsanyi:2016ksw, Petreczky:2016vrs}. The former was temperature-interpolated according to the DIGA prediction $\chi^{1/4} \sim T^{-2}$ and the isospin-breaking factor was removed. The latter was mass-extrapolated according to $\chi^{1/4}\sim m_\pi$. Right: behavior of $c_{\SP}/c_{\gluo}$ as a function of $M/m_s$ within the $M$-range. A straight horizontal line is set at $0$.}
\label{fig:systematic_chi_t_430}
\end{figure}

Finally, in Fig.~\ref{fig:systematic_chi_t_430}, we show how the continuum
limit of $\chi_{\SP}$ varies as a function of $M/m_s$ within the $M$-range
interval determined before (left plot). Also at finite temperature we observe that the continuum limit of $\chi_{\SP}^{1/4}$ is quite stable within the $M$-range; nevertheless the
residual systematic is incorporated in our final result, also shown in Fig.~\ref{fig:systematic_chi_t_430}. As for the gluonic case, the final error was instead estimated from the systematic variation of the continuum extrapolation varying the fit function and the fit range, similarly to the procedure followed at zero temperature. Finally, we also checked that performing a common fit to our gluonic results and our SP determinations for a fixed value of $M/m_s$ gave consistent results with the extrapolations performed for the SP determinations alone at the same value of $M/m_s$. As already done at zero temperature, we consider the conservative estimate $\chi^{1/4}_{\SP}=15(5)$~MeV as our final determination, where this error takes into account both the systematic and the statistical sources of uncertainty.

In Fig.~\ref{fig:systematic_chi_t_430}, in the right plot, we also report how the ratio $c_{\SP}/c_{\gluo}$ depends on $M/m_s$. As already observed when discussing the $T=0$ results, when $M/m_s$ is chosen small enough, the SP discretization suffers for much smaller lattice artifacts compared to the gluonic one. As this cut-off increases, the ratio $c_\SP/c_\gluo$ tends to grow, approaching $1$, which can be interpreted as the effect of the inclusion of more and more non-chiral modes in the spectral sums.

We now want to compare our final SP result for $\chi^{1/4}$ at $T\simeq430~\text{MeV}$, $\chi^{1/4} = 15(5)~\text{MeV}$, with other determinations at this temperature. This
value is compatible with our gluonic estimation ($\chi^{1/4}=14(5)~\text{MeV}$)
and is also compatible within $\sim 1.2\,\sigma$ with the value that can be
obtained interpolating results of Ref.~\cite{Borsanyi:2016ksw} in Tab.~S9 according to the DIGA prediction $\chi^{1/4} \sim T^{-2}$ and removing the isospin-breaking factor $0.88$ the authors use to restore the $u-d$ mass difference in their $N_f=2+1+1$ results for the susceptibility:
$\chi^{1/4}=9.0(4)$~MeV. Our determination is also in agreement within the errors with the gluonic result reported in Ref.~\cite{Petreczky:2016vrs} for this temperature: $\chi^{1/4}=11(2)$~MeV. Note however that in this case the comparison is less strict as the latter result was obtained rescaling gluonic determinations for $\chi^{1/4}$ reported in Ref.~\cite{Petreczky:2016vrs} by $(135~\text{MeV})/m_\pi$, as they were obtained for $m_\pi \approx 160$~MeV. Of course this procedure assumes that $\chi$ depends on the light quark mass as predicted by the DIGA: $\chi^{1/4} \sim m_l^{1/2} \sim m_\pi$. Finally, we mention that in Ref.~\cite{Petreczky:2016vrs} the authors also computed $\chi^{1/4}$ adopting a fermionic definition based on the disconnected chiral susceptibility, which was however fully consistent with the gluonic results used above.

\subsection{$\chi(T)$ from spectral projectors and comparison with the DIGA}\label{sec:power_law_behaviour} 

In our $N_f=2+1$ setup, semiclassical and perturbative approximations predict for the topological susceptibility when $T\gg T_c$ the scaling (see
Eq.~\eqref{eq:DIGA_prediction_chi})
\beq\label{eq:power_law_behaviour}
\chi^{1/4}(T)\propto T^{-b},\qquad b_{\DIGA}=2.
\eeq
The aim of the present section is to study the behavior of our results for
$\chi^{1/4}$ as a function of $T$ and to compare it with the DIGA prediction. 
Our final results for $\chi^{1/4}$ as a function of $T/T_c$, obtained both from the
gluonic and the SP definitions, are reported in
Tab.~\ref{tab:final_values_chi_vs_T}. These results were obtained following exactly the same strategy outlined in Sec.~\ref{sec:results_t_430} for $T\simeq 430$~MeV, and more details about them can be found in Appendix~\ref{sec:appendix_finite_T_res}.
\begin{table}[!htb]
\begin{center}
\begin{tabular}{ |c|c||c|c|}
\hline
$T$~[MeV] & $T/T_c$ & $\chi_{\SP}^{1/4}$~[MeV] & $\chi_{\gluo}^{1/4}$~[MeV] \\
\hline
230 & 1.48 & 49(11)    & 38(8) \\
300 & 1.94 & 41(8)     & 32(10) \\
365 & 2.35 & 26.5(5.5) & 20(7) \\
430 & 2.77 & 15(5)     & 14(5) \\
570 & 3.68 & 8(6)      & 6(6) \\
\hline
\end{tabular}
\end{center}
\caption{Results for the fourth root of the topological susceptibility as a function of $T$. For the crossover temperature $T_c$ we adopted the reference value $T_c=155$~MeV.}
\label{tab:final_values_chi_vs_T}
\end{table}

\begin{figure}[!t]
\centering
\includegraphics[scale=0.44]{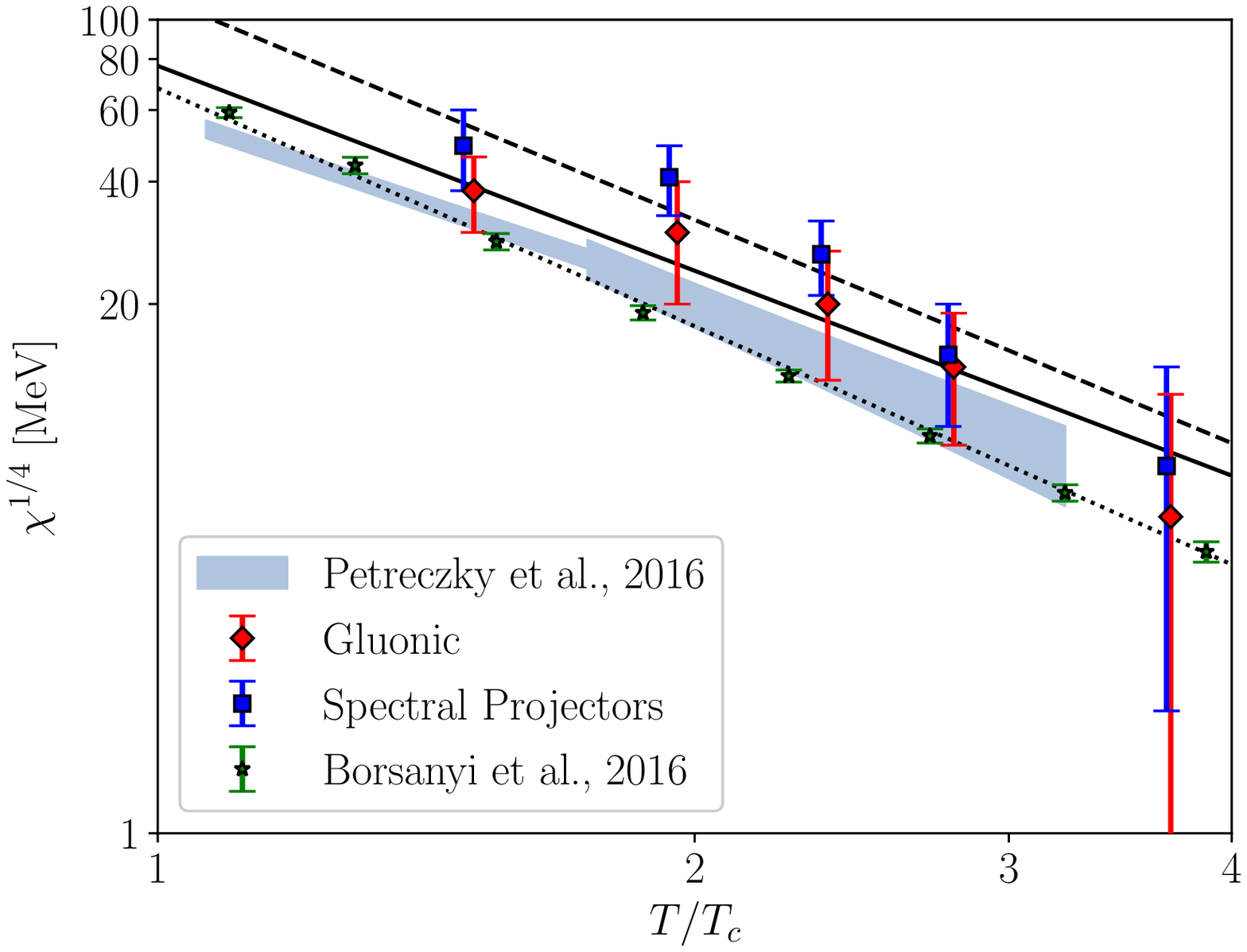}
\includegraphics[scale=0.44]{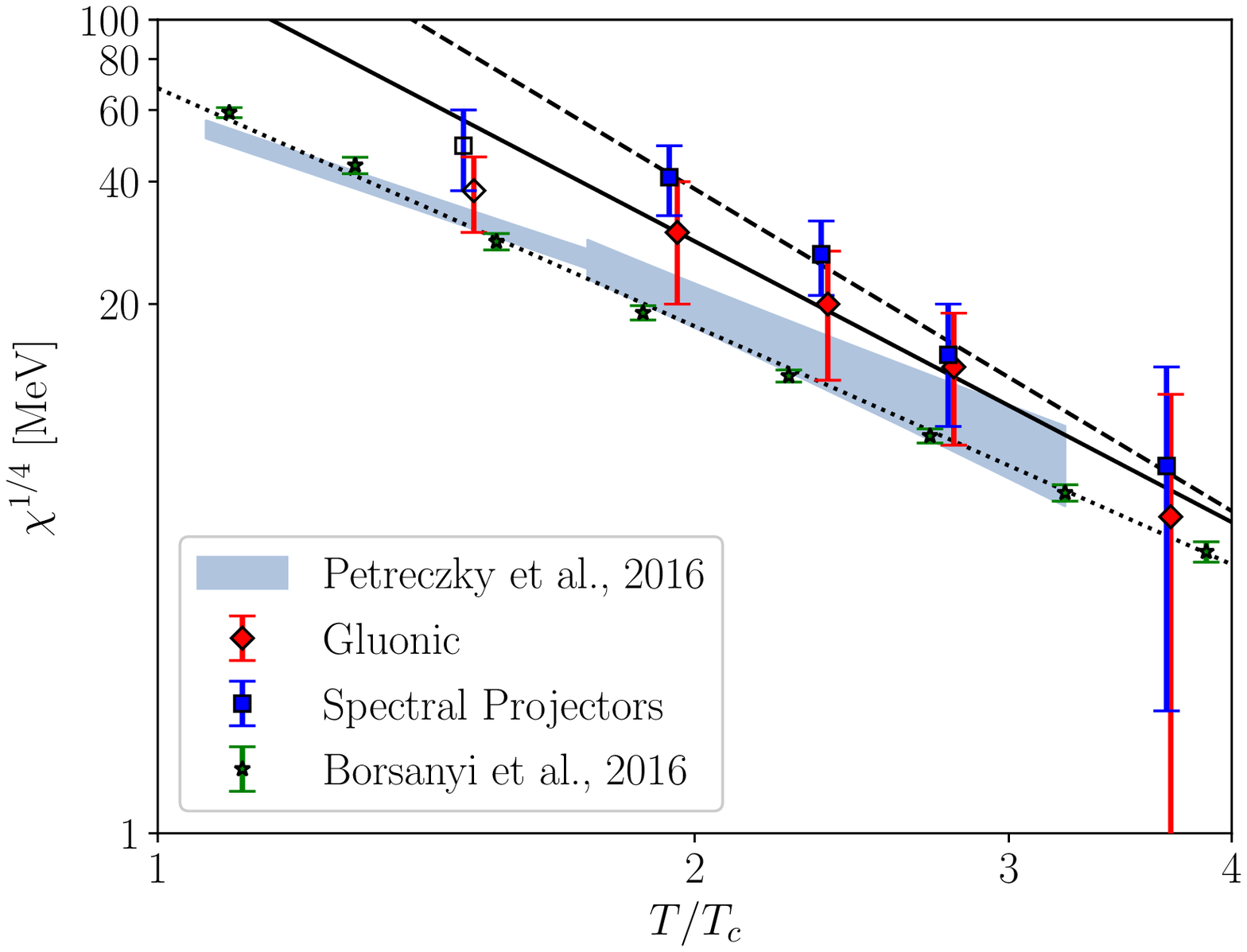}
\caption{Behavior of $\chi^{1/4}$ as a function of $T/T_c$ in double-$\log$ scale for our SP and gluonic data. Gluonic points are slightly shifted to improve readability. Starred points represent results taken from Ref.~\cite{Borsanyi:2016ksw}, where the isospin-breaking factor was removed, while the uniform-shaded area represents the gluonic determinations reported in Ref.~\cite{Petreczky:2016vrs}, which have been mass-rescaled according to $\chi^{1/4}\sim m_\pi$. The left plot reports the best fits of our data according to the fit function~\eqref{eq:fit_function_chi_vs_T} and performed including all available points: $\tilde{\chi}^2/\dof = 0.54/3$ for gluonic data and $\tilde{\chi}^2/\dof = 2.1/3$ for SP ones. The right plot reports the result of the same fits, but performed excluding our lowest temperature $T\simeq 230$~MeV (excluded points are empty). In this case, $\tilde{\chi}^2/\dof = 0.06/2$ for gluonic data and $\tilde{\chi}^2/\dof = 0.22/2$ for SP ones. Dashed, solid and dotted lines represent, respectively, best fits of our gluonic data, of our spectral data and of gluonic data of Ref.~\cite{Borsanyi:2016ksw}.}
\label{fig:chi_vs_T}
\end{figure}

In Fig.~\ref{fig:chi_vs_T}, we show the behavior of $\chi^{1/4}$ as a function
of $T/T_c$, from which it should be clear that our data are compatible with a power-law behavior of the type~\ref{eq:power_law_behaviour} in the whole explored range. If we perform a best fit of the data using the function
\beq\label{eq:fit_function_chi_vs_T}
\chi^{1/4} = A \left(\frac{T}{T_c}\right)^{-b}
\eeq
when including in the fit all available points, we obtain the results
\beqnn
b_{\SP}=1.82(43), \qquad b_{\gluo}=1.67(51)\ .
\eeqnn
If we instead exclude from the fit the lowest temperature $T=230$~MeV, we get:
\beqnn
b_{\SP}=2.63(81), \qquad b_{\gluo}=2.3(1.1)\ .
\eeqnn
We thus conclude that our data are in perfect agreement with a power-law
behavior in the whole explored range, with an effective exponent that turns out
to be well compatible with $b_{\DIGA}$ within our errors already for
$T\gtrsim 300$~MeV, i.e., already for $T/T_c \gtrsim 2$. Actually, also when $T=230$~MeV is included in the fit the exponent $b_\SP$ turns out to be compatible within $1$ standard deviation with $b_\DIGA$. However, as it can be clearly seen from Figs.~\ref{fig:chi_vs_T}, the inclusion/exclusion of such point visibly changes the slope of the fit. This may be an indication, albeit at present not conclusive, that the effective exponent $b$ changes when going from $T\sim 200$~MeV to $T\sim300$~MeV.

In Fig.~\ref{fig:chi_vs_T} we also compare our results with previous determinations reported in Refs.~\cite{Borsanyi:2016ksw,Petreczky:2016vrs}. Concerning results of Ref.~\cite{Borsanyi:2016ksw}, to make a fair comparison, we removed the isospin-breaking factor the authors use to restore the $u-d$ mass difference in their $N_f=2+1+1$ results. Our spectral determinations are systematically larger than those of Ref.~\cite{Borsanyi:2016ksw}, and we find a $\sim2.5-3$ standard deviation tension at $T=300$~MeV and $T=365$~MeV. As for the temperature behavior, we observe that the best fit of the data of Ref.~\cite{Borsanyi:2016ksw} for $T\gtrsim 170$~MeV with the fit
function~\eqref{eq:fit_function_chi_vs_T} yields $b=1.945(23)$, i.e., in this case the DIGA-like power-law seems to set in for smaller values of the temperature.

Concerning the gluonic results of Ref.~\cite{Petreczky:2016vrs}, since they were obtained for $m_\pi = 160$~MeV, we rescaled them according to the DIGA prediction $\chi^{1/4}\sim m_\pi$ in order to make a fair comparison. Also in this case we observe that our determinations lay systematically above. Although this comparison is less strict because of the different pion masses adopted, also in this case we observe a $\sim 2-2.5$ standard deviation tension among determinations for $T=300$~MeV and $T=365$~MeV.

The observed tensions between our spectral determinations and previous results reported in the literature deserve to be further investigated, for example by refining the present errors on the spectral determinations for $T\lesssim 400$~MeV. Also probing higher temperatures with our methods would be interesting in order to extend the present comparison towards the $\sim 1$~GeV region, which is also interesting in the context of axion cosmology.

\section{Conclusions}\label{sec:conclusions}

In this work we performed a numerical lattice study of the behavior of the
topological susceptibility $\chi(T)$ in the high temperature regime of QCD with
$N_f=2+1$ quarks at the physical point. 

Our computational strategy relies on the discretization of the topological
charge through Spectral Projectors on the eigenmodes of the staggered Dirac
operator. The reason for this choice is to reduce the large lattice artifacts
affecting the gluonic definition of $\chi$ at high temperatures when non-chiral
quarks, like the staggered ones, are adopted to discretize the QCD action. The
problem of the dominance of the $Q=0$ sector, which is related to the
suppression of $\chi$ at high-$T$, is instead addressed by the use of the
multicanonical method already applied for this purpose in the recent
work~\cite{Bonati:2018blm}.

The spectral definition of the topological susceptibility introduces a new free
parameter, as the sum over the chiralities of the eigenmodes of the lattice
staggered Dirac operator is performed by including all eigenvalues with
magnitude $\vert \lambda \vert \le M$.  In principle, any value of $M$ (kept
constant in physical units as $a\to 0$) would provide a correct continuum limit of
$\chi$; however, since isolating WBZMs is highly ambiguous, we identified a
reasonable range of $M$-values on which to perform the continuum limit at fixed
$M$. Any residual systematic related to the choice of $M$ is then assessed
afterwards and included in our final determination of the uncertainty affecting
$\chi$.

To test the spectral method we first of all study the $T=0$ case, where $\chi$
can be reliably computed by using chiral perturbation theory.  In this case the
multicanonical algorithm is not needed, since the MC evolution naturally visits
$Q\neq 0$ sectors, and the final result obtained by the spectral method is
perfectly compatible both with the NLO ChPT result and with
previous gluonic estimates. 

The advantage of the spectral projectors method with respect to the gluonic one
is that lattice artifacts are in this case much smaller, and the extrapolation
towards the continuum limit is thus better under control.  Moreover, due to the
presence of the parameter $M$, in the spectral projectors setting we have a
natural procedure to check for the presence of systematics in the continuum
extrapolation procedure: to look for residual dependence on $M$ of the
extrapolated result.  As a matter of fact, this systematic gives the dominant contribution to the final error on $\chi$ for some of the temperatures studied in this work.

In the high temperature regime, we explored 5 values of $T$, ranging from
$\sim200$~MeV to $\sim600$~MeV. Also in these cases we observe good agreement
between SP and gluonic data, with the spectral results that are generically
more accurate than the gluonic ones; however, more important is that, as
noted above, in the SP case we have a much better control of the systematics of
the continuum extrapolation.

We finally investigated the behavior of our results for $\chi$ as a function of
$T$, comparing with expectations based on the DIGA approximation. We find that a
decaying power law well describes our data in the whole explored range; in
particular the effective exponent of this power law is well in agreement with
the DIGA prediction for $T\gtrsim 300$~MeV, i.e., for $T/T_c \gtrsim 2$. This
is in agreement with the results obtained in Ref.~\cite{Petreczky:2016vrs} and
is consistent with a growing number of observations suggesting that
high-temperature QCD is dominated by strong non-perturbative effects for
temperatures going approximately from the chiral crossover up to
$\sim300$~MeV~\cite{Alexandru:2019gdm,Alexandru:2021pap,Kotov:2021rah,Cardinali:2021mfh}.

We remark that our results for $\chi^{1/4}$ from spectral projectors show a
$\sim2-3$ standard deviation tension in the range $300~\mathrm{MeV}\lesssim T
\lesssim 400~\mathrm{MeV}$ range when compared with previous determinations in Refs.~\cite{Borsanyi:2016ksw,Petreczky:2016vrs}. Moreover, also when the
consistency between our results and the ones in the literature is better, our
spectral determinations for $\chi^{1/4}(T)$ systematically points to larger
values in the whole explored range. The same behavior, when observed in the
gluonic determinations of $\chi$, could be ascribed to a problem of the
continuum extrapolation, that is unable to capture the asymptotic $O(a^2)$
scaling and introduces a bias in the extrapolation. The lattice spacing
dependence of the SP determinations is however much milder than that of the
gluonic estimates, and such an interpretation of the observed disagreement seems
unlikely in this case. In conclusion, the picture emerging from the comparison carried out
in Fig.~\ref{fig:chi_vs_T} is that a complete quantitative understanding of the
behavior of $\chi(T)$ in the high temperature regime of QCD is still missing,
and further studies will be required to clarify the sources of the observed
tensions between different determinations.

Several directions can be followed to improve the present study: first of all
it seems crucial to refine our present estimates of the topological
susceptibility, in order to make the comparison with the results of
Refs.~\cite{Borsanyi:2016ksw, Petreczky:2016vrs} more stringent, and obtain a
precise and unbiased estimate of $\chi(T)$ in the explored temperature range.
For this purpose simulations with larger statistics and smaller lattice
spacings are required. Other natural extensions include a systematic study of
the temperature ranges $T\lesssim 400$~MeV, where deviations from the DIGA
power law should be visible, and $T\sim 1$~GeV, in order to reach the region of
temperatures directly relevant for axion cosmology.

However, simulations at smaller lattice spacings (needed both to improve the
present estimates and to reach higher temperatures) are practically unfeasible
with standard RHMC simulations, due to the severe topological critical slowing
down. A promising strategy to overcome this problem is the \textit{parallel
tempering on boundary conditions} algorithm proposed in
Ref.~\cite{Hasenbusch:2017unr}, which has already been successfully applied
both to two dimensional models~\cite{Hasenbusch:2017unr, Berni:2019bch} and to
$4d$ $\mathrm{SU}(N)$ Yang-Mills theories without matter
fields~\cite{Bonanno:2020hht,Bonanno:2022yjr}. 

\section*{Acknowledgements}

A.~A.~has been financially supported by the European Union’s Horizon 2020 research and innovation programme ``Tips in SCQFT'' under the Marie Sk\l{}odowska-Curie grant agreement No.~791122, as well as by the Horizon 2020 European research infrastructures programme ``NI4OS-Europe'' with grant agreement no.~857645. C.~Bonanno acknowledges the support of the Italian Ministry of Education, University and Research under the project PRIN 2017E44HRF, ``Low dimensional quantum systems: theory, experiments and simulations''. M.~D.~thanks Sayantan Sharma for useful discussions. Numerical simulations have been performed on the \texttt{MARCONI} and \texttt{MARCONI100} machines at CINECA, based on the agreement between INFN and CINECA (under projects INF19\_npqcd, INF20\_npqcd and INF21\_npqcd).

\appendix

\section*{Appendix}
\section{Summary of finite temperature results for $\chi$}\label{sec:appendix_finite_T_res}

We here report all the results for the topological susceptibility $\chi$ at finite temperature not shown in Sec.~\ref{sec:results_t_430}, following the same order of presentation of that section. Each subsection refers to a single temperature and results are shown as follows:
\begin{itemize}
\item scatter plot of the chiralities of the lowest eigenmodes of the staggered operator for the finest lattice spacing explored at that temperature;
\item comparison of the continuum limits obtained for the gluonic and the SP definitions;
\item assessment of the systematics related to the choice of $M/m_s$.
\end{itemize}

The parameters of all the performed simulations are summarized in Tab.~\ref{tab:simulation_parameters_all_T_remaining}. For the two coarsest lattice spacings simulated at $T\simeq 230$~MeV, no enhancement of topological fluctuations was needed. Therefore, simulations of these points were carried on without the multicanonic algorithm.

\begin{table}[!htb]
\begin{center}
\begin{tabular}{ |c|c||c|c|c|c|c|}
\hline
$T$~[MeV] & $T/T_c$ & $\beta$ & $a$~[fm] & $\hat{m}_s \cdot 10^{-2}$ & $N_s$ & $N_t$ \\
\hline
\hline
\multirow{5}{*}{230} & \multirow{5}{*}{1.48} & 3.814* & 0.1073 & 4.27 & 32 & 8  \\
&& 3.918* & 0.0857 & 3.43 & 40 & 10 \\
&& 4.014  & 0.0715 & 2.83 & 48 & 12 \\
&& 4.100  & 0.0613 & 2.40 & 56 & 14 \\
&& 4.181  & 0.0536 & 2.10 & 64 & 16 \\
\hline	
\hline		
\multirow{4}{*}{300} & \multirow{4}{*}{1.94} & 3.938 & 0.0824 & 3.30 & 32 & 8  \\
&& 4.059  & 0.0659 & 2.60 & 40 & 10 \\
&& 4.165  & 0.0549 & 2.15 & 48 & 12 \\
&& 4.263  & 0.0470 & 1.86 & 56 & 14 \\
\hline
\hline		
\multirow{4}{*}{365} & \multirow{4}{*}{2.35} & 4.045 & 0.0676 & 2.66 & 32 & 8 \\
&& 4.175  & 0.0541 & 2.12 & 40 & 10 \\
&& 4.288  & 0.0451 & 1.78 & 48 & 12 \\
&& 4.377  & 0.0386 & 1.55 & 56 & 14 \\
\hline
\hline
\multirow{4}{*}{570} & \multirow{4}{*}{3.68} & 4.140 & 0.0572 & 2.24 & 24 & 6 \\
&& 4.316  & 0.0429 & 1.71 & 32 & 8 \\
&& 4.459  & 0.0343 & 1.37 & 40 & 10 \\
&& 4.592  & 0.0286 & 1.09 & 48 & 12 \\
\hline
\end{tabular}
\end{center}
\caption{Simulation parameters used for $T\simeq 230$~MeV, $T\simeq 300$~MeV,
$T\simeq 365$~MeV and $T\simeq 570$~MeV. The bare parameters $\beta$,
$\hat{m}_s$ and the lattice spacings have been fixed according to results of
Refs.~\cite{Aoki:2009sc, Borsanyi:2010cj, Borsanyi:2013bia}, and $\hat{m}_{l}$
is fixed through $\hat{m}_s/\hat{m}_{l}=m_s/m_l=28.15$. Simulations marked with * have been performed without multicanonic algorithm.}
\label{tab:simulation_parameters_all_T_remaining}
\end{table}

\FloatBarrier
\newpage

\subsection{$T=230~\text{MeV}$}

\begin{figure}[!htb]
\centering
\includegraphics[scale=0.39]{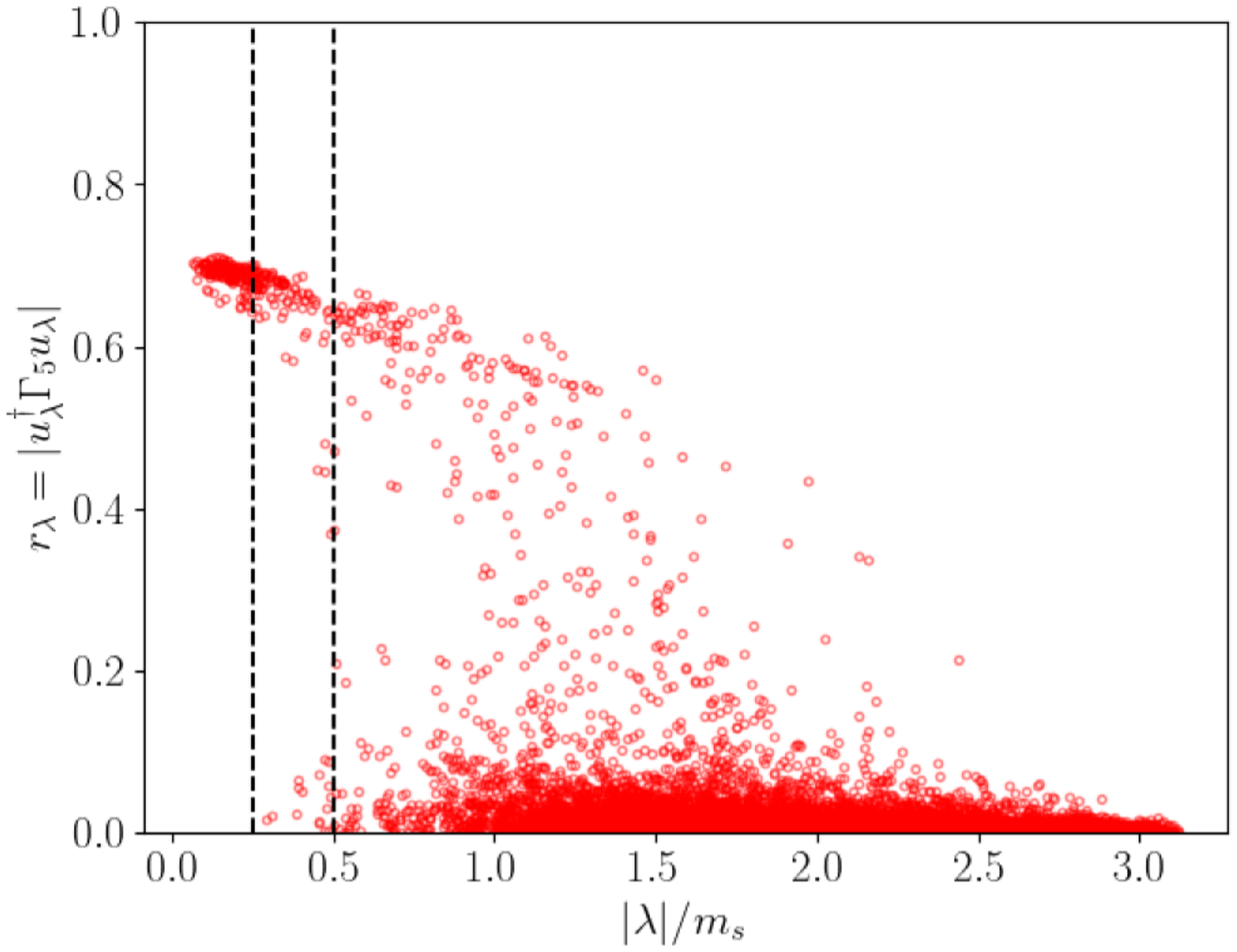}
\hspace{2mm}
\includegraphics[scale=0.39]{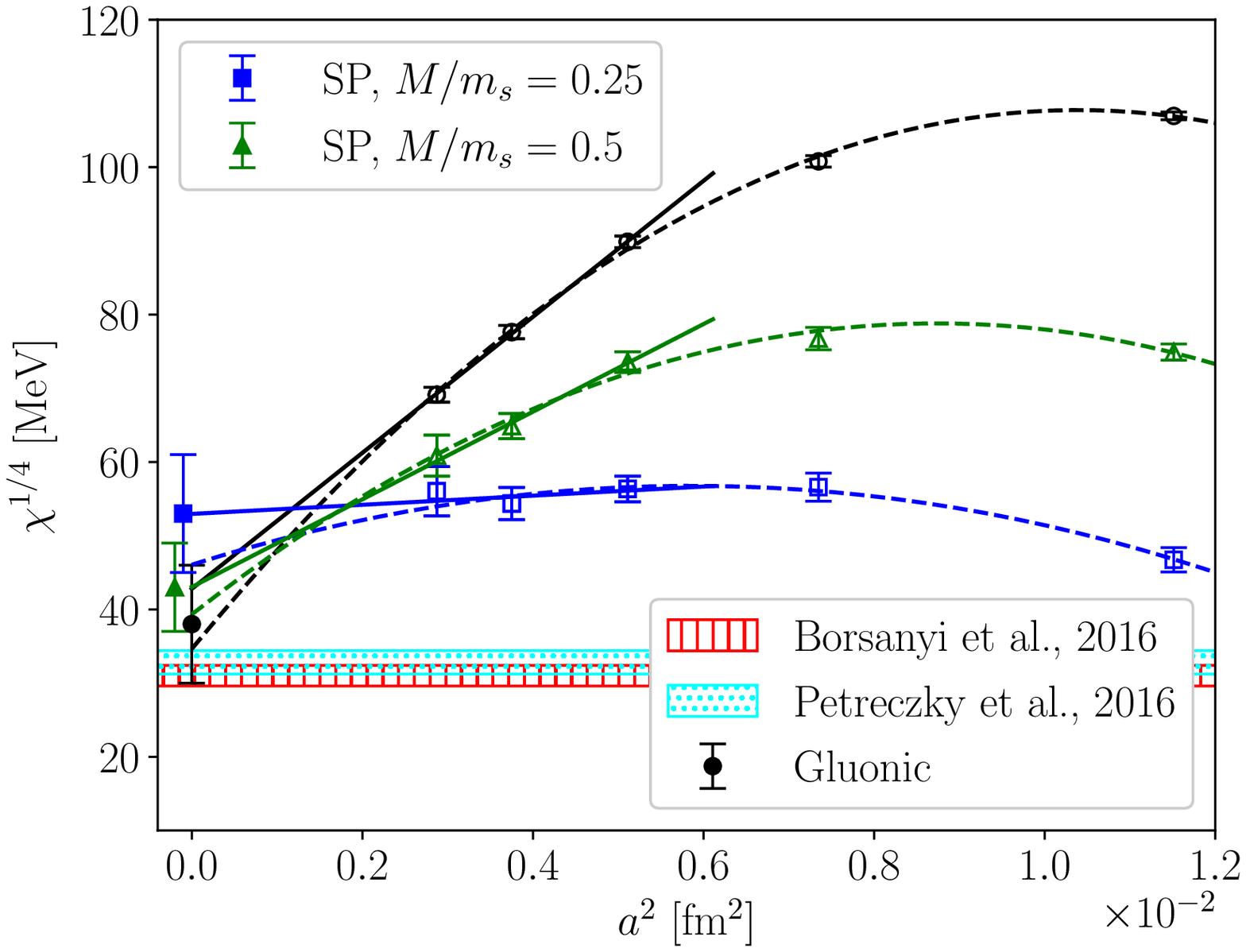}
\caption{Left: scatter plot of the chirality $r_\lambda$ vs
$\vert\lambda\vert/m_s$ for our run with $a\simeq0.0536$~fm at $T\simeq230~\text{MeV}$. The
two dashed vertical lines are set at $0.25$ and $0.5$ and delimit the chosen
interval for $M/m_s$. For each reported configuration, only the
first $200$ eigenvalues (with the lowest magnitude) are shown. Right: Comparison
of the continuum limits of $\chi^{1/4}$ for $T\simeq230$~MeV obtained with the
gluonic and the SP discretizations. Vertically-hatched and dotted-hatched bands display the values of $\chi^{1/4}$ obtained for this temperature in, respectively, Refs.~\cite{Borsanyi:2016ksw, Petreczky:2016vrs}. The former was temperature-interpolated according to the DIGA prediction $\chi^{1/4} \sim T^{-2}$ and the isospin-breaking factor was removed. The latter was mass-extrapolated according to $\chi^{1/4}\sim m_\pi$.}
\label{fig:scatter_plot_t_230}
\end{figure}

\begin{figure}[!htb]
\centering
\includegraphics[scale=0.39]{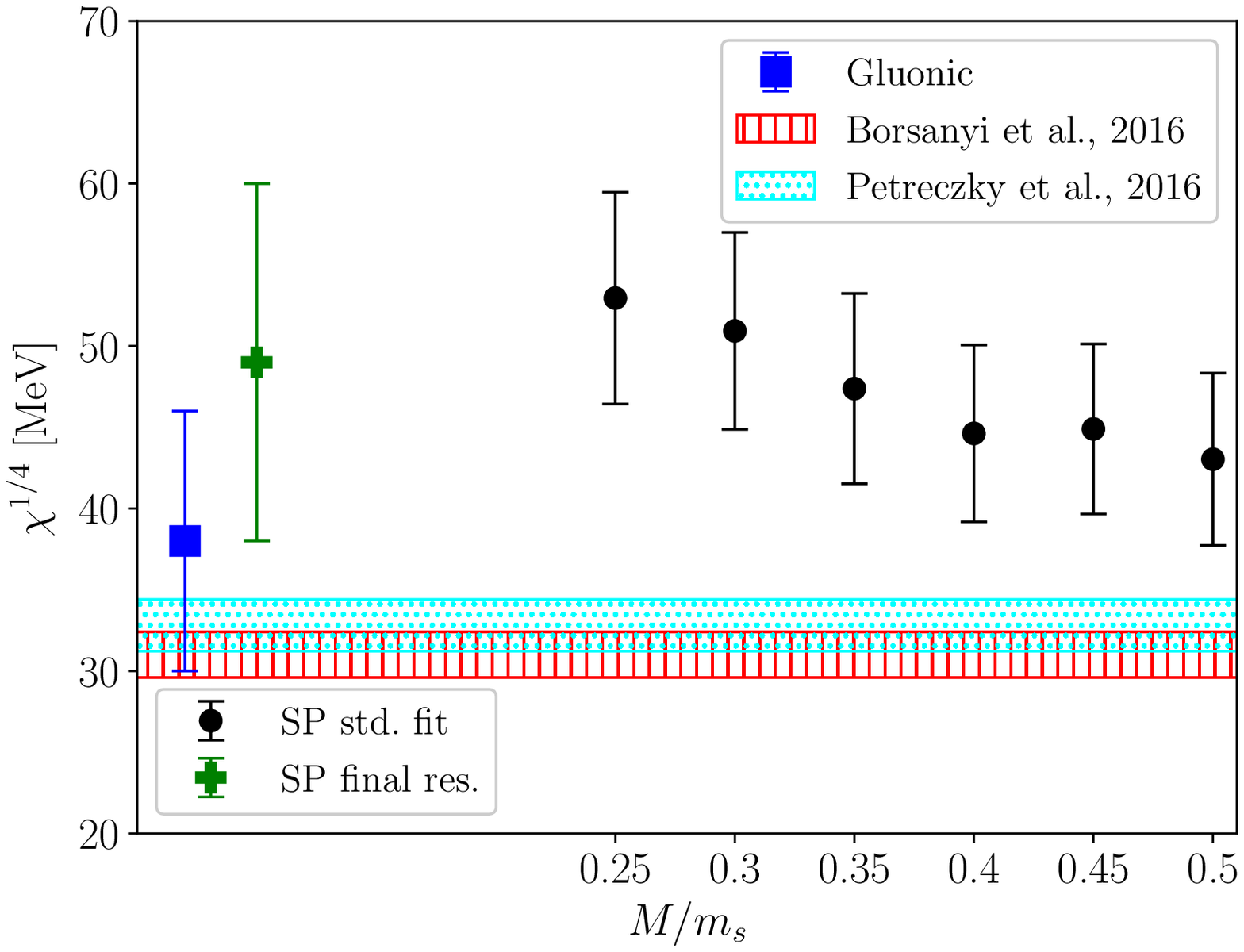}
\hspace{2mm}
\includegraphics[scale=0.39]{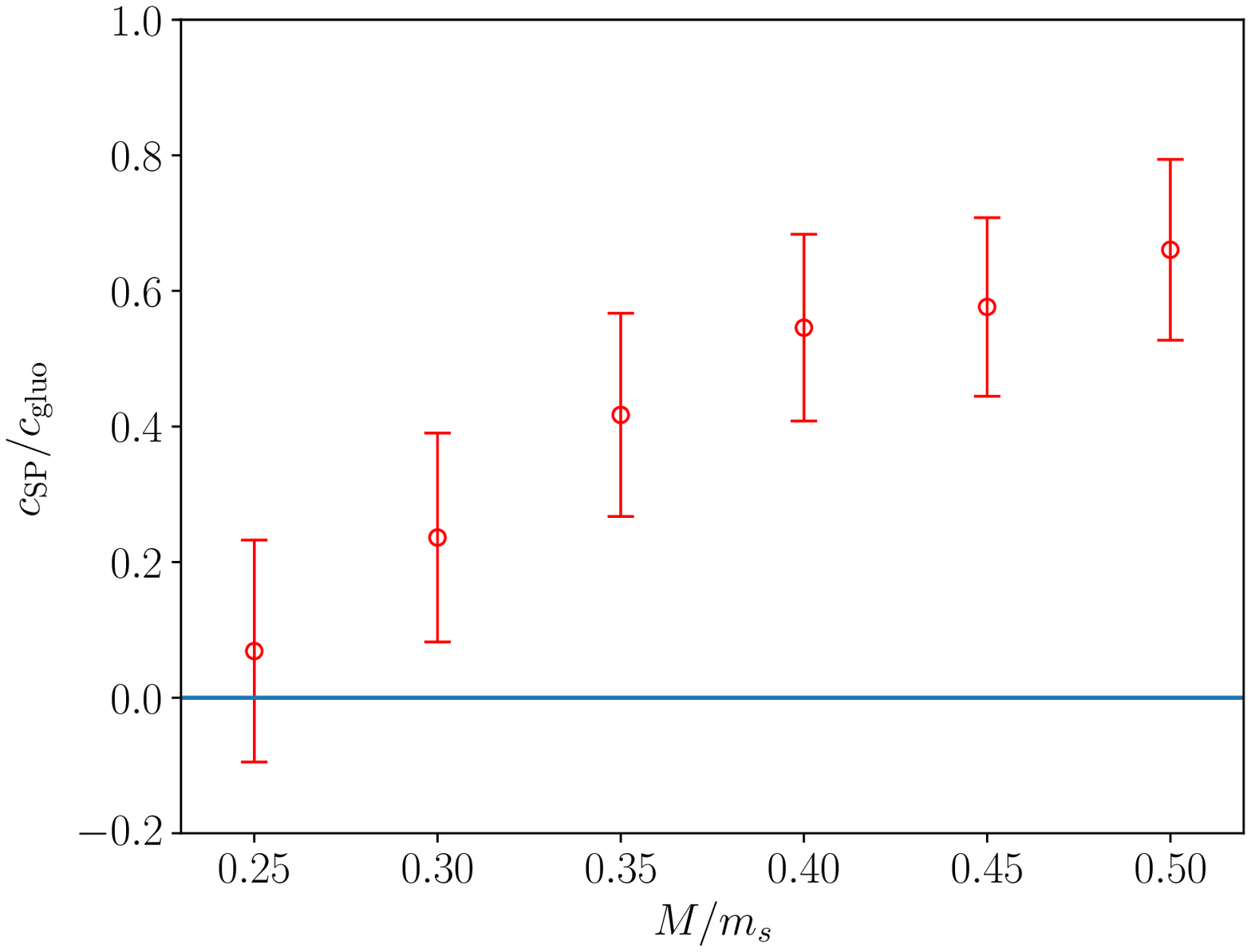}
\caption{Left: continuum limits of $\chi^{1/4}$ obtained at $T\simeq 230~\text{MeV}$ from spectral
projectors for several values of $M/m_s$ within the $M$-range. Each error bar refers to the continuum extrapolation obtained fitting the three finest lattice spacings with the fit function reported in Eq.~\eqref{eq:SP_cont_scaling}. The cross
point is our final SP result, which includes any residual systematic related to
the choice of $M/m_s$. Vertically-hatched and dotted-hatched bands display the values of $\chi^{1/4}$ obtained for this temperature in, respectively, Refs.~\cite{Borsanyi:2016ksw, Petreczky:2016vrs}. The former was temperature-interpolated according to the DIGA prediction $\chi^{1/4} \sim T^{-2}$ and the isospin-breaking factor was removed. The latter was mass-extrapolated according to $\chi^{1/4}\sim m_\pi$. Right: $c_{\SP}/c_{\gluo}$ for the several values of $M/m_s$ within the $M$-range. A straight horizontal line is set at $0$.}
\label{fig:systematic_chi_t_230}
\end{figure}

\FloatBarrier
\newpage

\subsection{$T=300~\text{MeV}$}

\begin{figure}[!htb]
\centering
\includegraphics[scale=0.39]{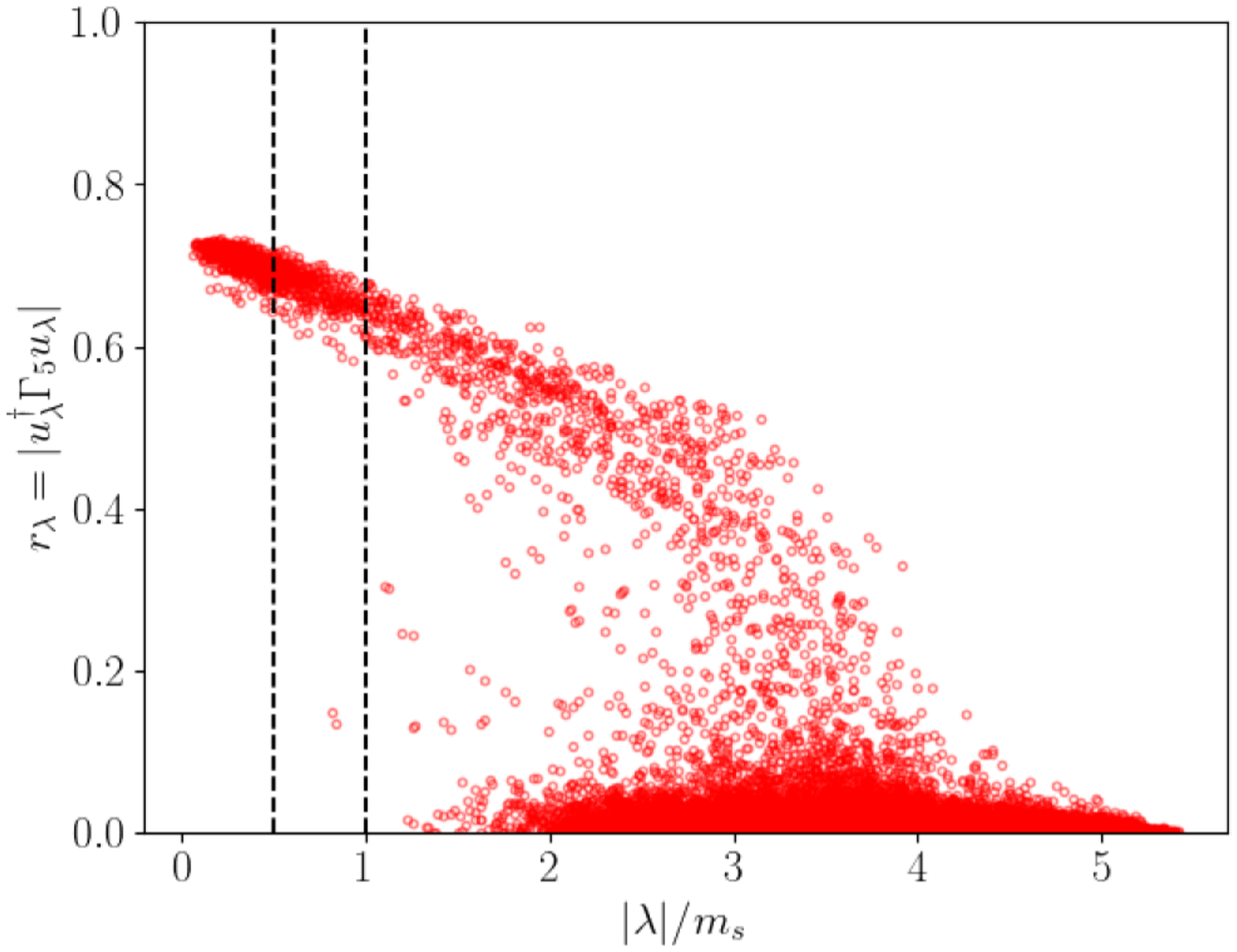}
\hspace{2mm}
\includegraphics[scale=0.39]{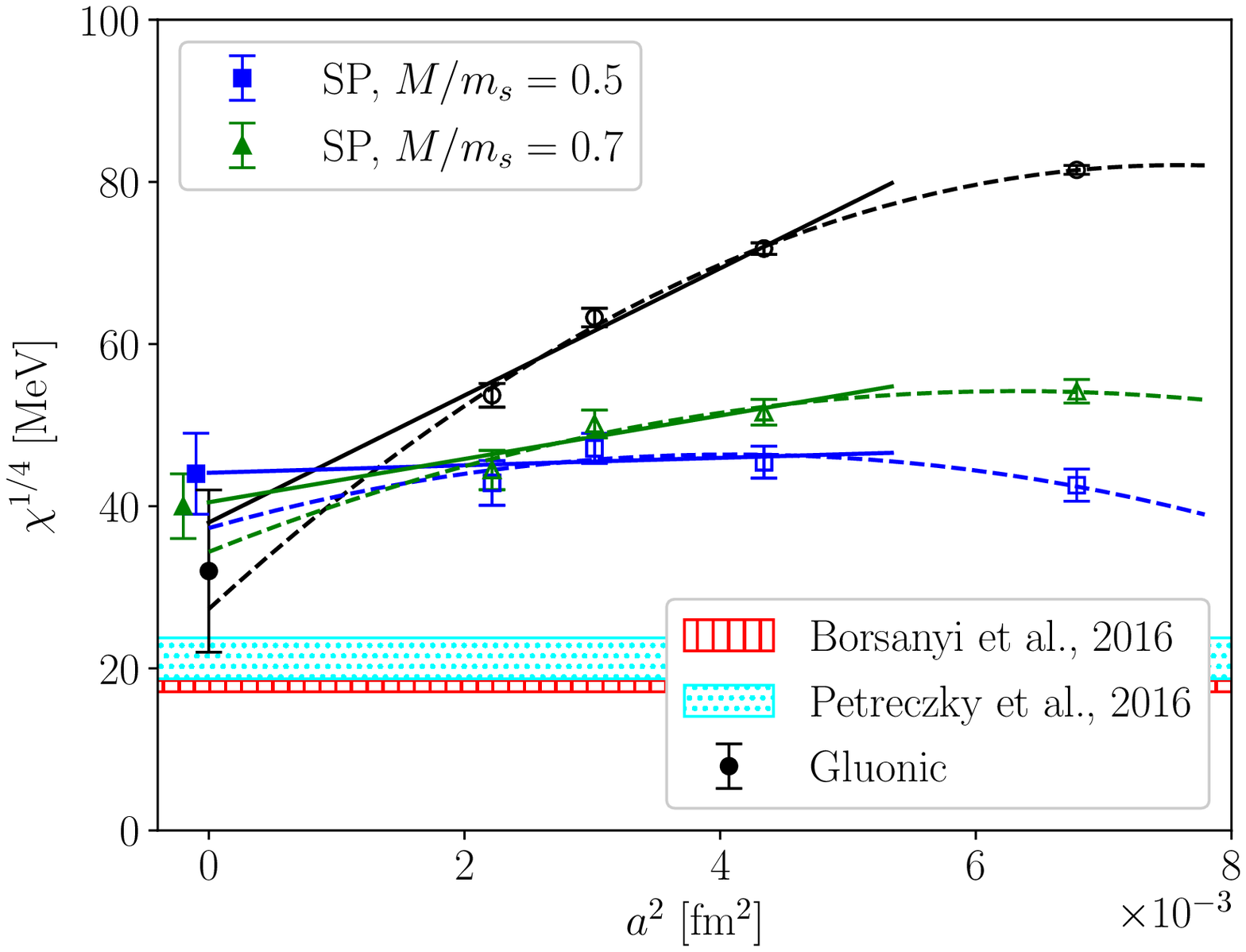}
\caption{Left: scatter plot of the chirality $r_\lambda$ vs
$\vert\lambda\vert/m_s$ for our run with $a\simeq0.0470$~fm at $T\simeq300~\text{MeV}$. The
two dashed vertical lines are set at $0.5$ and $1$ and delimit the chosen interval
for $M/m_s$. For each reported configuration, only the first
$200$ eigenvalues (with the lowest magnitude) are shown. Right: Comparison of
the continuum limits of $\chi^{1/4}$ for $T\simeq300$~MeV obtained with the
gluonic and the SP discretizations. Vertically-hatched and dotted-hatched bands display the values of $\chi^{1/4}$ obtained for this temperature in, respectively, Refs.~\cite{Borsanyi:2016ksw, Petreczky:2016vrs}. The former was temperature-interpolated according to the DIGA prediction $\chi^{1/4} \sim T^{-2}$ and the isospin-breaking factor was removed. The latter was mass-extrapolated according to $\chi^{1/4}\sim m_\pi$.}
\label{fig:scatter_plot_t_300}
\end{figure}

\begin{figure}[!htb]
\centering
\includegraphics[scale=0.39]{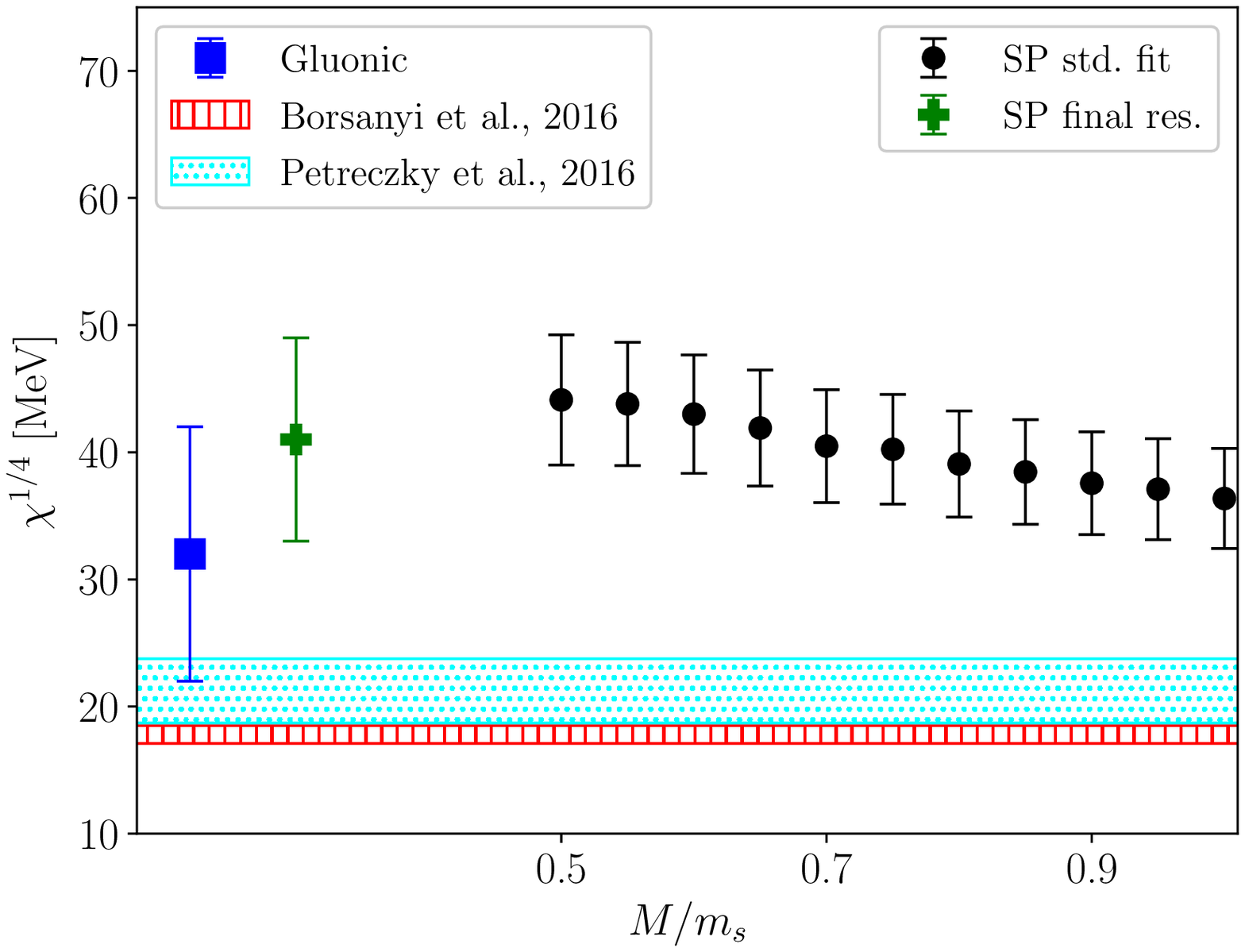}
\hspace{2mm}
\includegraphics[scale=0.39]{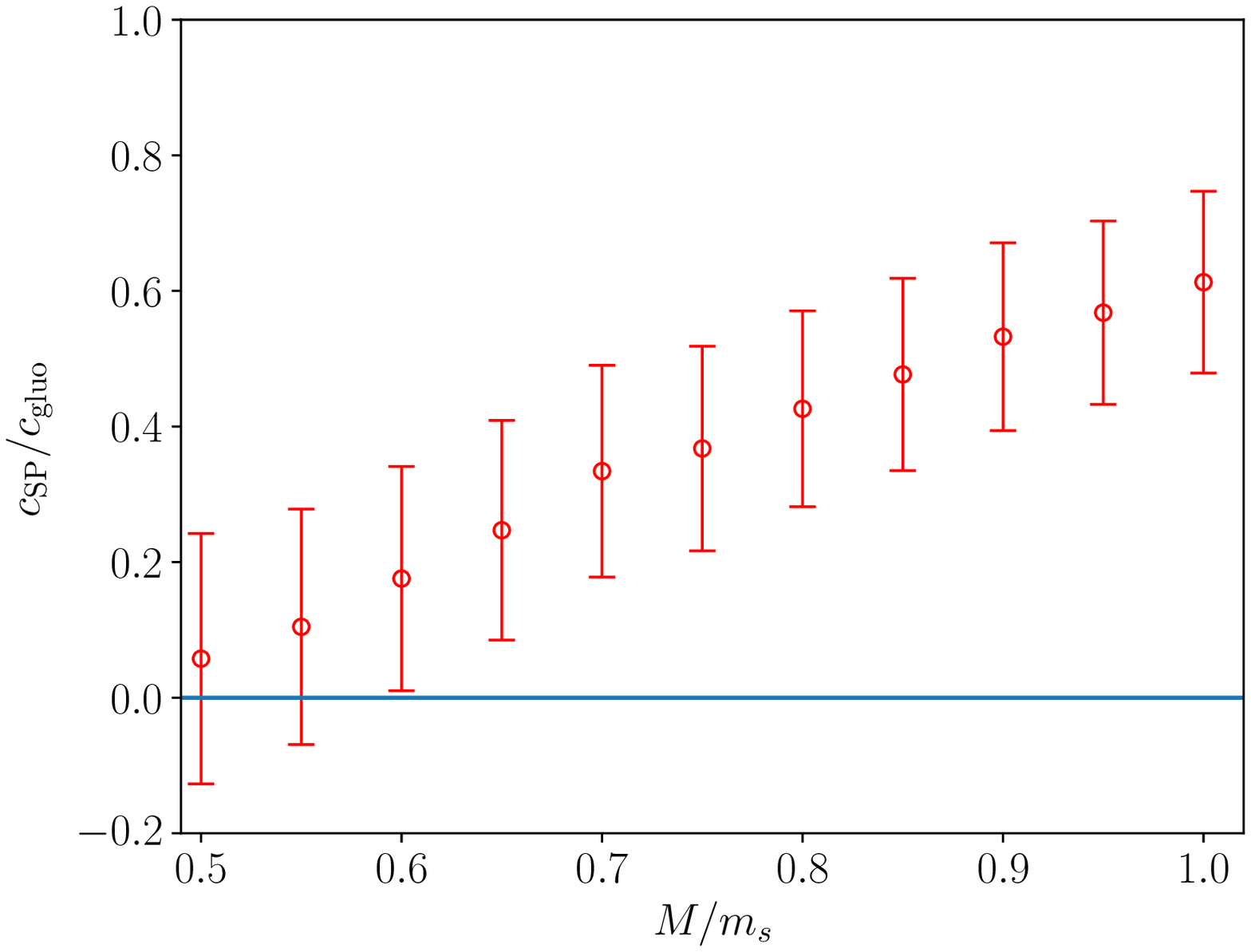}
\caption{Left: continuum limits of $\chi^{1/4}$ obtained at $T\simeq 300~\text{MeV}$ from spectral
projectors for several values of $M/m_s$ within the $M$-range. Each error
bar refers to the continuum extrapolation obtained fitting the three finest lattice spacings with the fit function reported in Eq.~\eqref{eq:SP_cont_scaling}. The cross
point is our final SP result, which includes any residual systematic related to
the choice of $M/m_s$. Vertically-hatched and dotted-hatched bands display the values of $\chi^{1/4}$ obtained for this temperature in, respectively, Refs.~\cite{Borsanyi:2016ksw, Petreczky:2016vrs}. The former was temperature-interpolated according to the DIGA prediction $\chi^{1/4} \sim T^{-2}$ and the isospin-breaking factor was removed. The latter was mass-extrapolated according to $\chi^{1/4}\sim m_\pi$. Right:
$c_{\SP}/c_{\gluo}$ for the several values of $M/m_s$ within the $M$-range. A straight horizontal line is set at $0$.}
\label{fig:systematic_chi_t_300}
\end{figure}

\FloatBarrier
\newpage

\subsection{$T=365~\text{MeV}$}

\begin{figure}[!htb]
\centering
\includegraphics[scale=0.39]{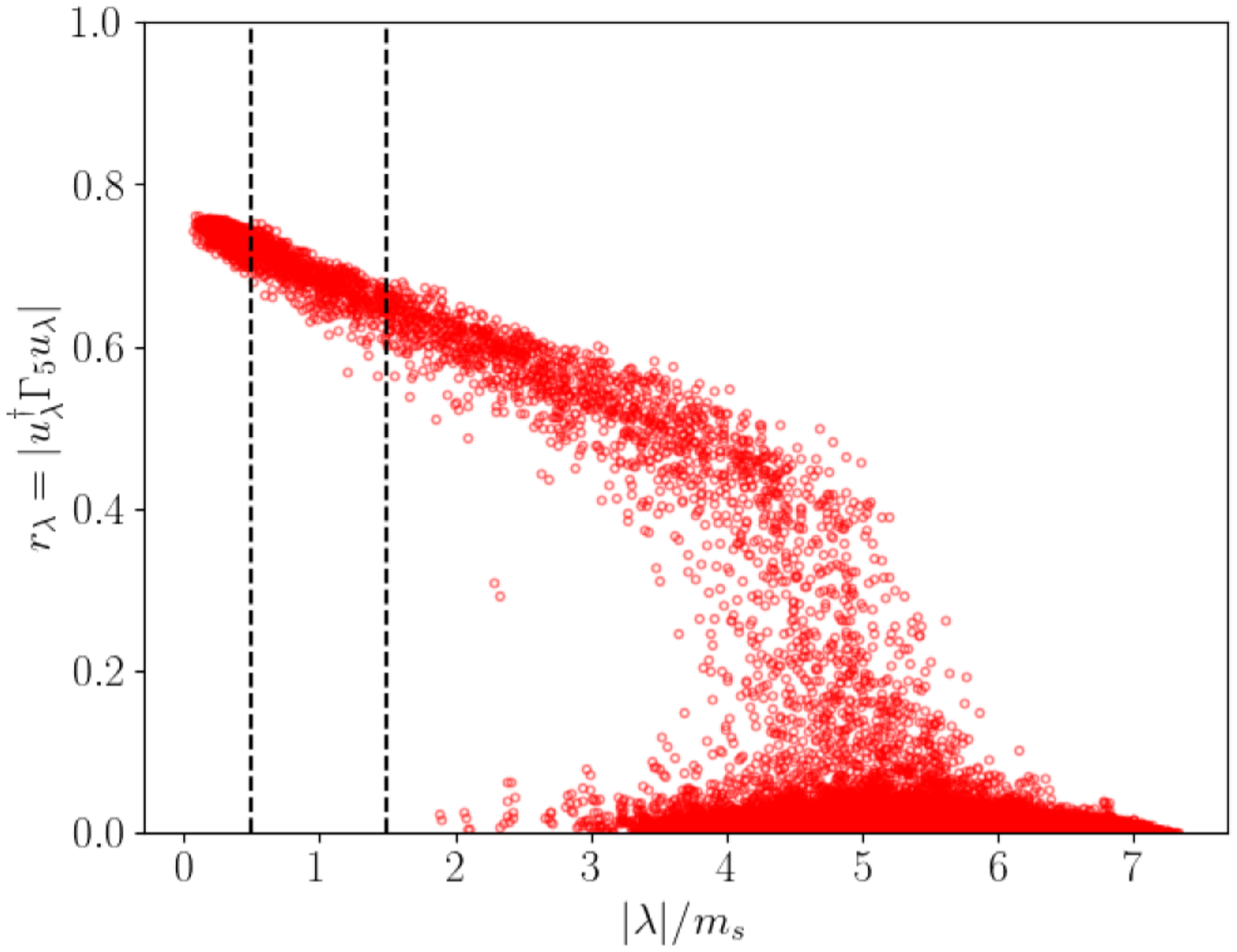}
\hspace{2mm}
\includegraphics[scale=0.39]{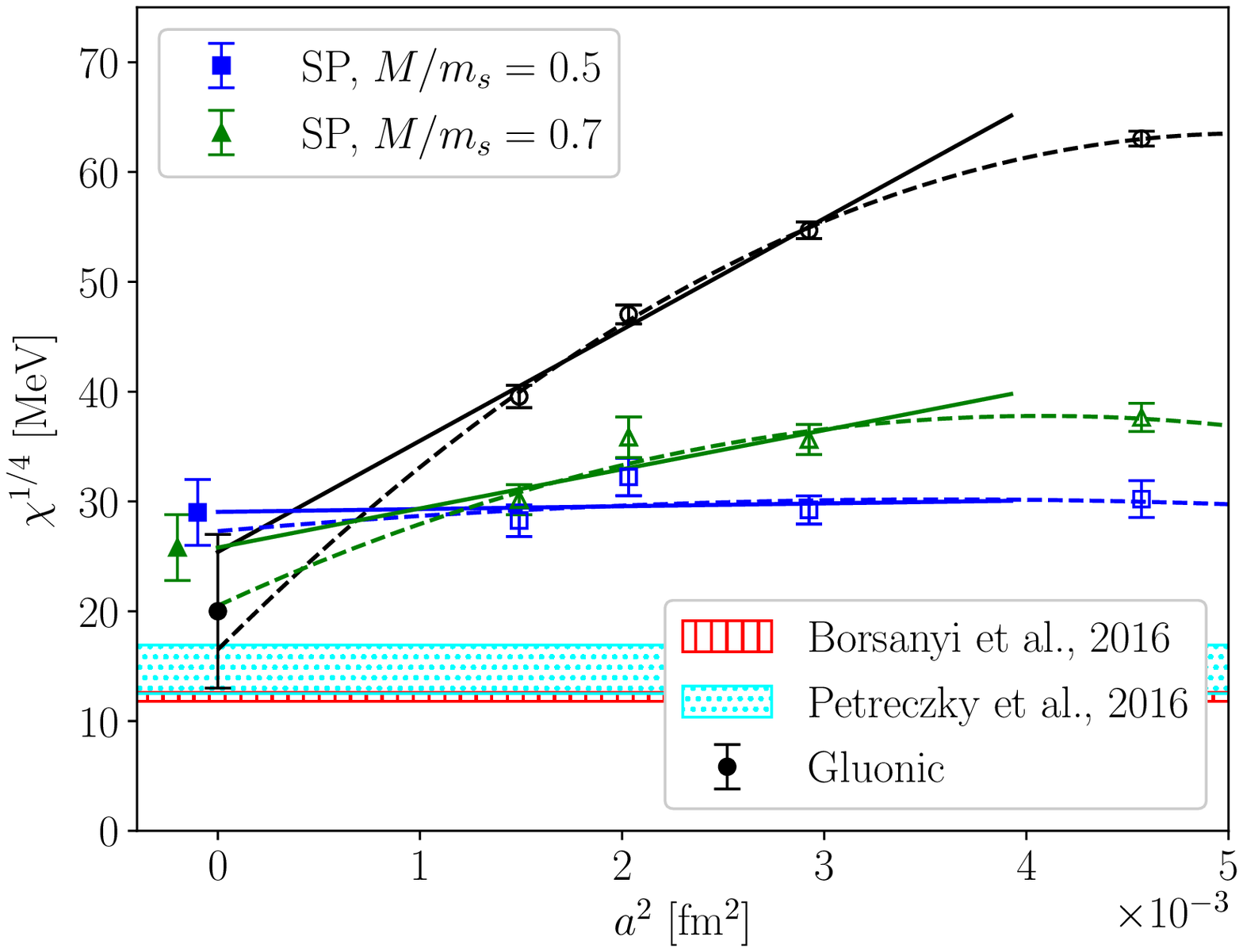}
\caption{Left: scatter plot of the chirality $r_\lambda$ vs
$\vert\lambda\vert/m_s$ for our run with $a\simeq0.0386$~fm at $T\simeq365~\text{MeV}$. The
two dashed vertical lines are set at $0.5$ and $1.5$ and delimit the chosen
interval for $M/m_s$. For each reported configuration, only the
first $200$ eigenvalues (with the lowest magnitude) are shown. Right: Comparison
of the continuum limits of $\chi^{1/4}$ for $T\simeq365$~MeV obtained with the
gluonic and the SP discretizations. Vertically-hatched and dotted-hatched bands display the values of $\chi^{1/4}$ obtained for this temperature in, respectively, Refs.~\cite{Borsanyi:2016ksw, Petreczky:2016vrs}. The former was temperature-interpolated according to the DIGA prediction $\chi^{1/4} \sim T^{-2}$ and the isospin-breaking factor was removed. The latter was mass-extrapolated according to $\chi^{1/4}\sim m_\pi$.}
\label{fig:scatter_plot_t_365}
\end{figure}
\begin{figure}[!htb]
\centering
\includegraphics[scale=0.39]{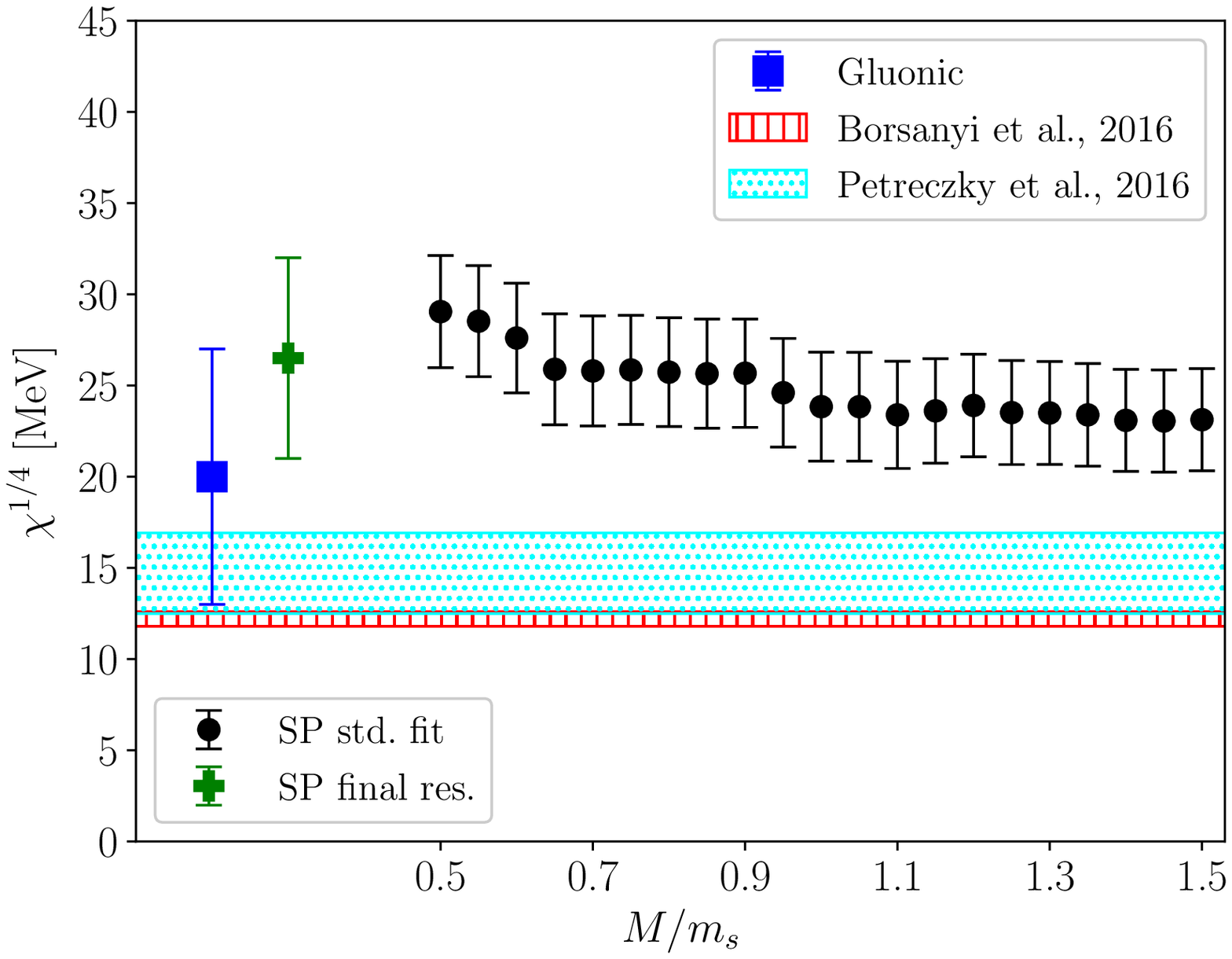}
\hspace{2mm}
\includegraphics[scale=0.39]{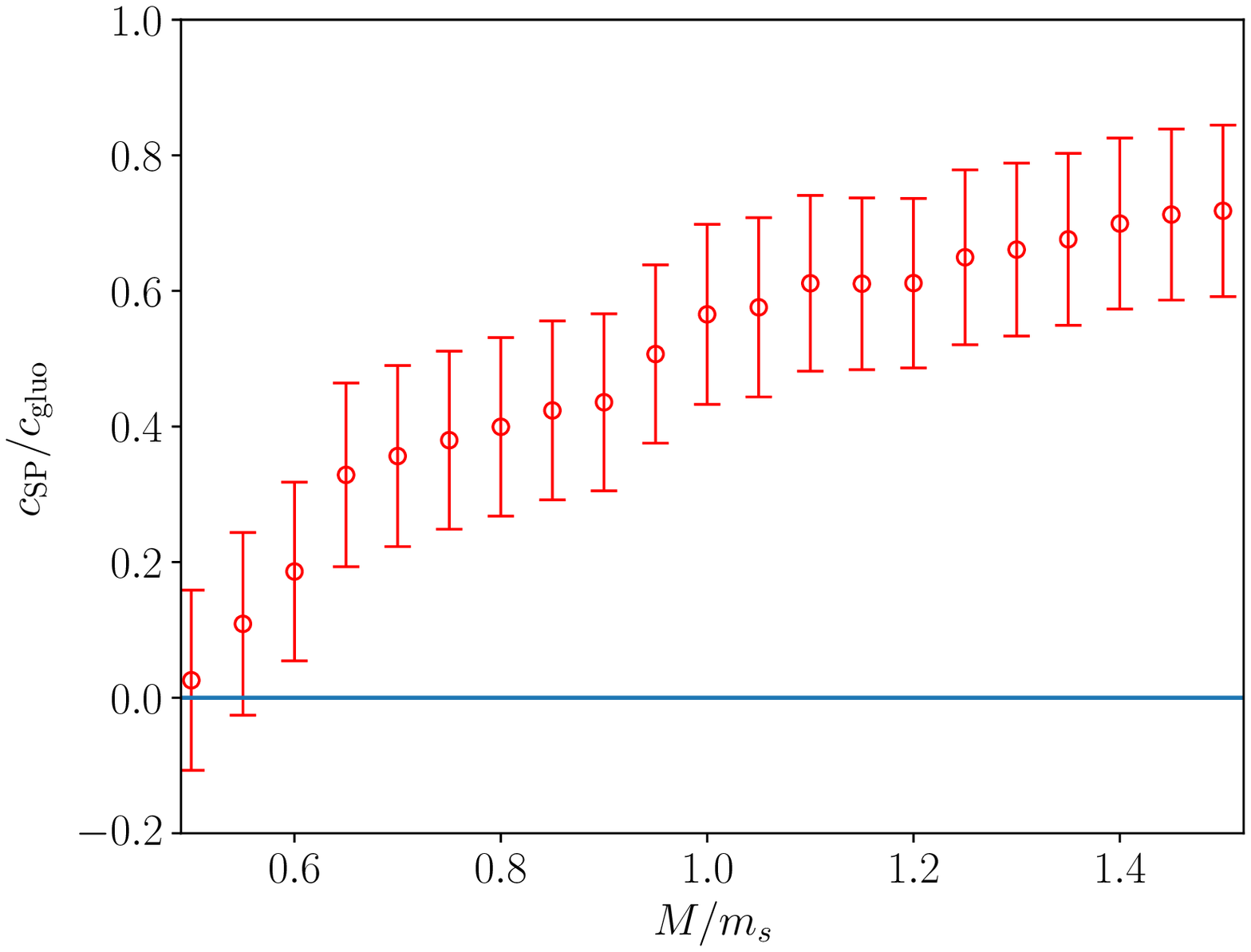}
\caption{Left: continuum limits of $\chi^{1/4}$ obtained at $T\simeq 365~\text{MeV}$ from spectral projectors for several values of $M/m_s$ within the $M$-range. Each error
bar refers to the continuum extrapolation obtained fitting the three finest lattice spacings with the fit function reported in Eq.~\eqref{eq:SP_cont_scaling}. The cross
point is our final SP result, which includes any residual systematic related to
the choice of $M/m_s$. Vertically-hatched and dotted-hatched bands display the values of $\chi^{1/4}$ obtained for this temperature in, respectively, Refs.~\cite{Borsanyi:2016ksw, Petreczky:2016vrs}. The former was temperature-interpolated according to the DIGA prediction $\chi^{1/4} \sim T^{-2}$ and the isospin-breaking factor was removed. The latter was mass-extrapolated according to $\chi^{1/4}\sim m_\pi$. Right: $c_{\SP}/c_{\gluo}$ for the several values of $M/m_s$ within the $M$-range. A straight horizontal line is set at $0$.}
\label{fig:systematic_chi_t_365}
\end{figure}

\FloatBarrier
\newpage

\subsection{$T=570~\text{MeV}$}
\begin{figure}[!htb]
\centering
\includegraphics[scale=0.39]{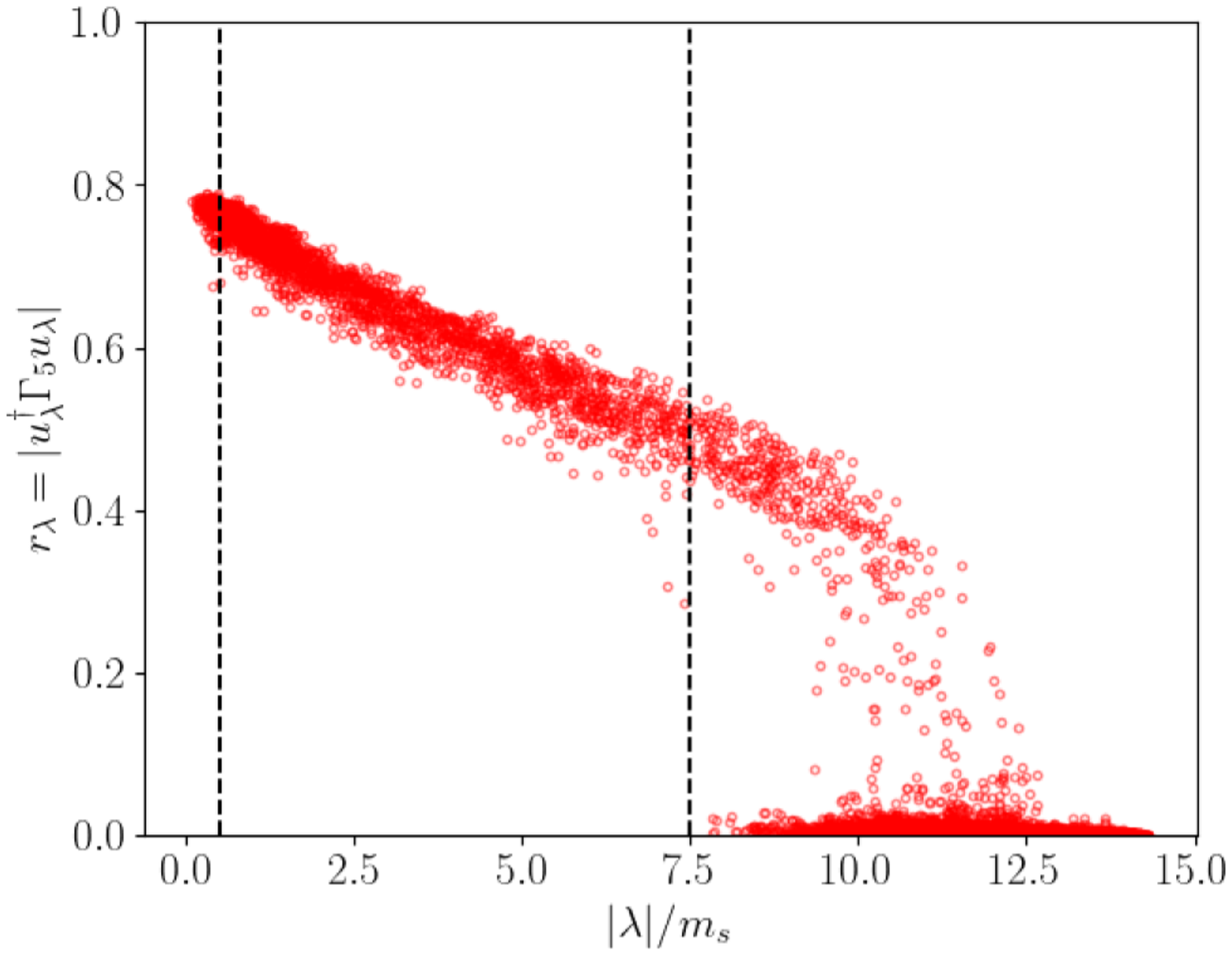}
\hspace{2mm}
\includegraphics[scale=0.39]{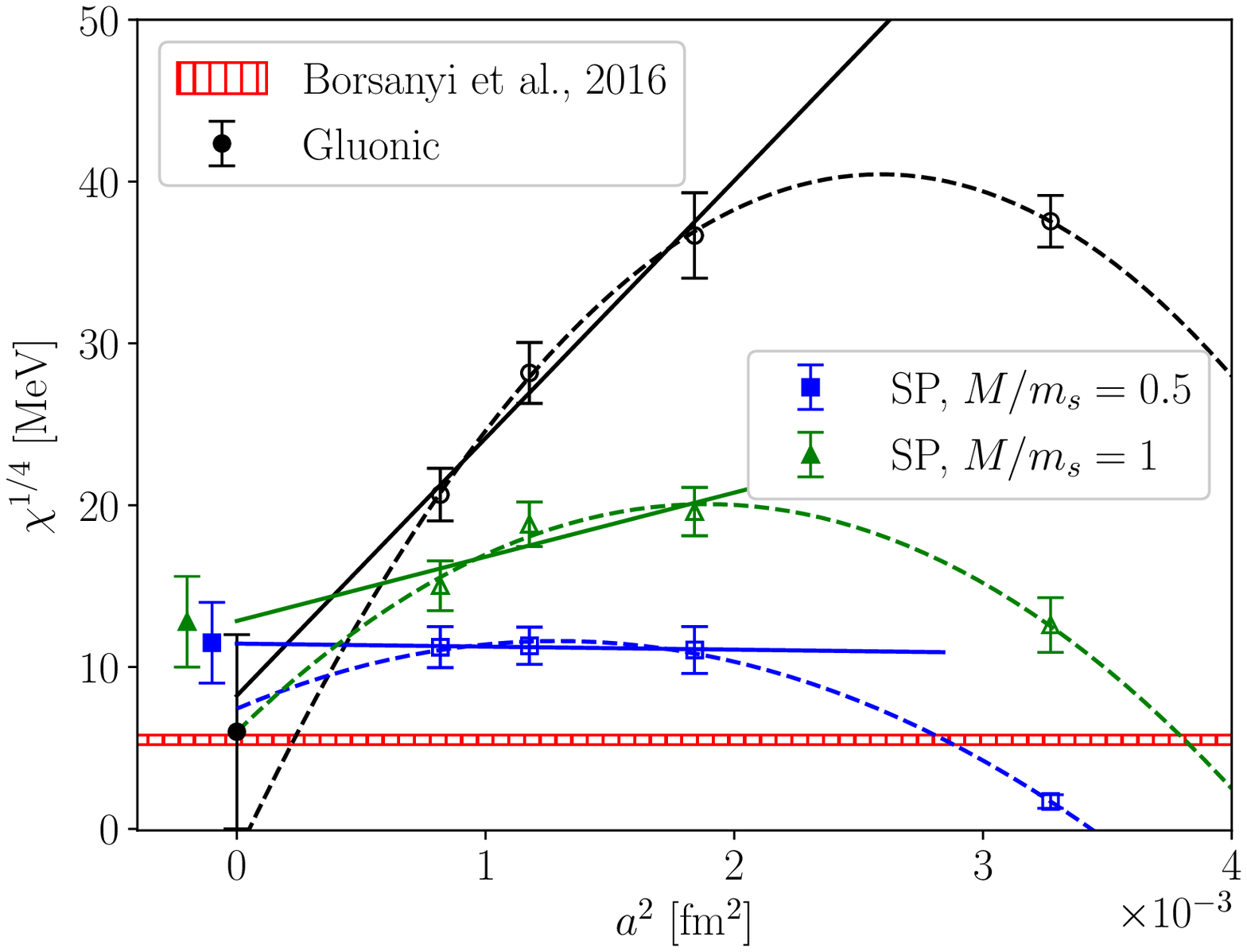}
\caption{Left: scatter plot of the chirality $r_\lambda$ vs
$\vert\lambda\vert/m_s$ for our run with $a\simeq0.0286$~fm at $T\simeq570~\text{MeV}$. The
two dashed vertical lines are set at $0.5$ and $7.5$ and delimit the chosen
interval for $M/m_s$. For each reported configuration, only the
first $200$ eigenvalues (with the lowest magnitude) are shown. Right: Comparison
of the continuum limits of $\chi^{1/4}$ for $T\simeq570$~MeV obtained with the
gluonic and the SP discretizations. Vertically-hatched band displays the value of $\chi^{1/4}$ obtained for this temperature in Ref.~\cite{Borsanyi:2016ksw}, which was temperature-interpolated according to the DIGA prediction $\chi^{1/4} \sim T^{-2}$ and the isospin-breaking factor was removed.}
\label{fig:scatter_plot_t_570}
\end{figure}

\begin{figure}[!htb]
\centering
\includegraphics[scale=0.39]{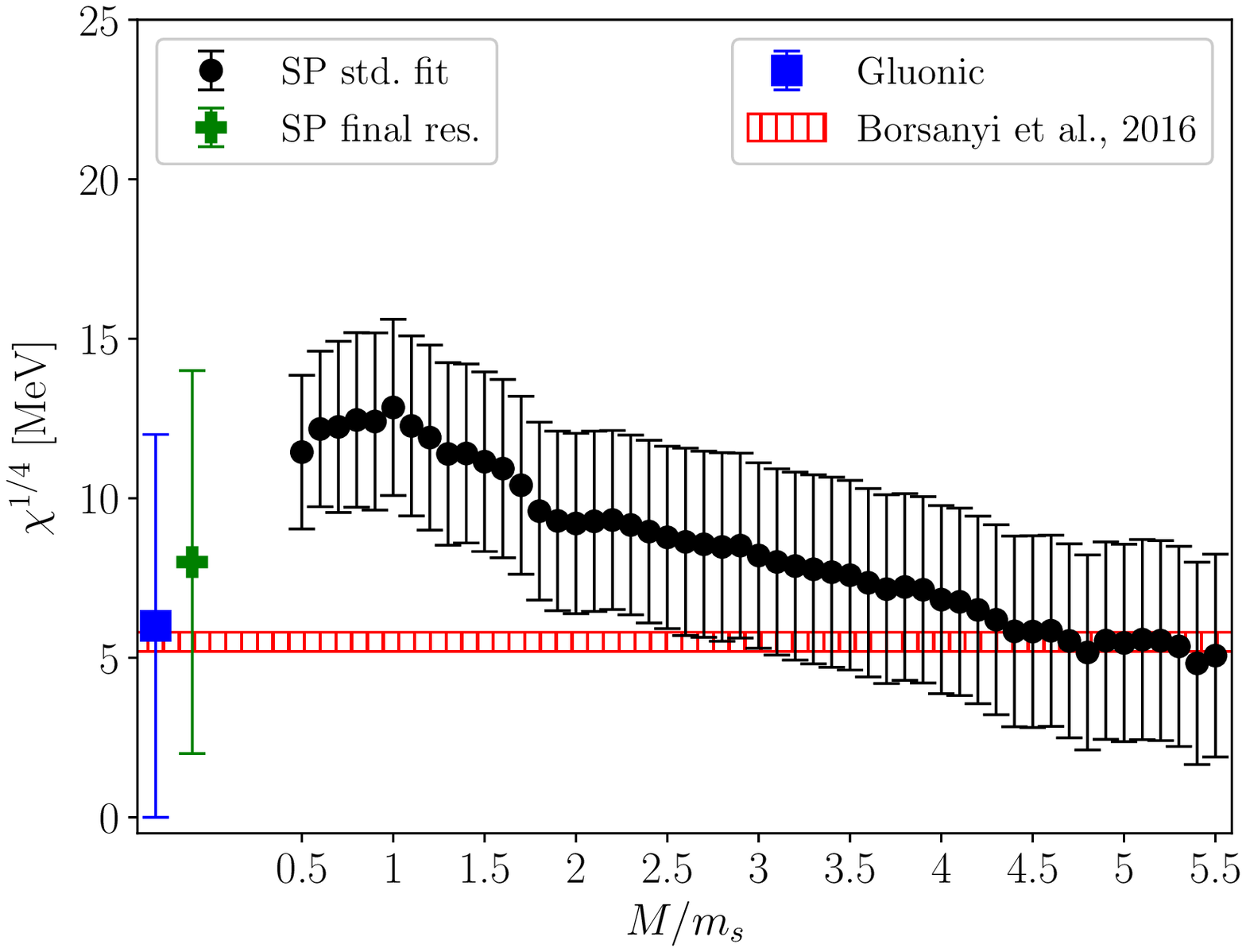}
\hspace{2mm}
\includegraphics[scale=0.39]{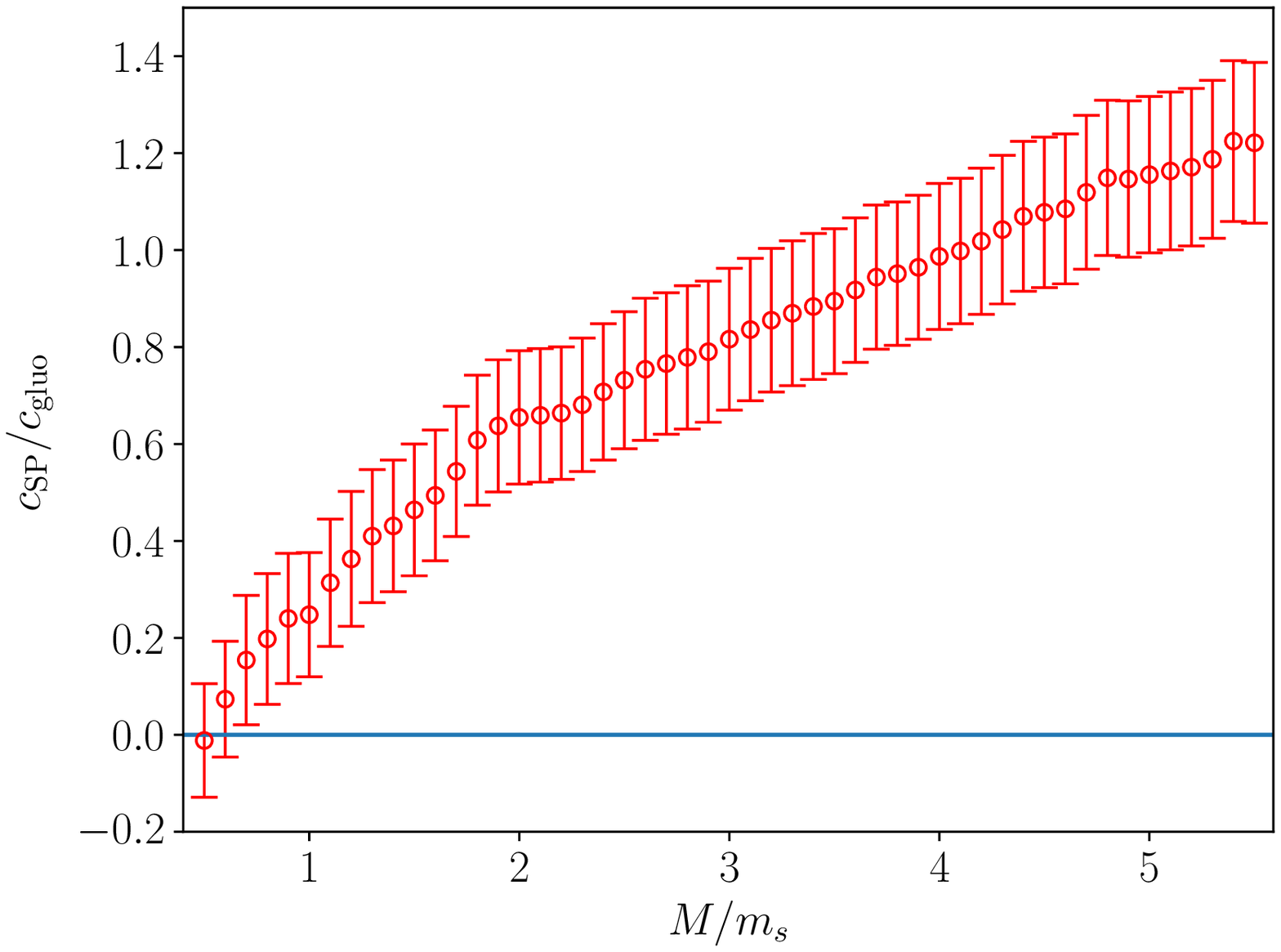}
\caption{Left: continuum limits of $\chi^{1/4}$ obtained at $T\simeq 570~\text{MeV}$ from spectral projectors for several values of $M/m_s$ within the $M$-range. Each error bar refers to the continuum extrapolation obtained fitting the three finest lattice spacings with the fit function reported in Eq.~\eqref{eq:SP_cont_scaling}. The cross
point is our final SP result, which includes any residual systematic related to
the choice of $M/m_s$. Vertically-hatched band displays the value of $\chi^{1/4}$ obtained for this temperature in Ref.~\cite{Borsanyi:2016ksw}, which was temperature-interpolated according to the DIGA prediction $\chi^{1/4} \sim T^{-2}$ and the isospin-breaking factor was removed. Right: $c_{\SP}/c_{\gluo}$ for the several values of $M/m_s$ within the $M$-range. A straight horizontal line is set at $0$.}
\label{fig:systematic_chi_t_570}
\end{figure}

The shown $M$-range for $\chi_\SP$ is narrower compared to the one shown in the scatter plot of chiralities, cf.~Figs.~\ref{fig:scatter_plot_t_570} and~\ref{fig:systematic_chi_t_570}, because the mean maximum eigenvalue of $D_\stag$ with periodic boundaries along the time direction was smaller than the anti-periodic case.

\FloatBarrier
\newpage

\providecommand{\href}[2]{#2}\begingroup\raggedright\endgroup

\end{document}